\documentclass[12pt,preprint]{aastex}

\shorttitle{New High Proper Motion Discoveries}
\shortauthors{Finch}

\begin{document}

\title{THE SOLAR NEIGHBORHOOD. XVIII. \\ DISCOVERY OF NEW PROPER
  MOTION STARS \\ WITH 0$\farcs$40 yr$^{-1}$ $>$ $\mu$ $\ge$
  0$\farcs$18 yr$^{-1}$ BETWEEN DECLINATIONS $-$90$\degr$ and
  $-$47$\degr$}

\author{Charlie T. Finch, Todd J. Henry, John P. Subasavage, and
  Wei-Chun Jao}

\affil{Georgia State University, Atlanta, GA 30302--4106}

\author{Nigel C. Hambly}

\affil{Scottish Universities Physics Alliance (SUPA), Institute for
  Astronomy, University of Edinburgh \\ Royal Observatory, Blackford
  Hill, Edinburgh, EH9~3HJ, Scotland, UK}

\email{finch@chara.gsu.edu}


\begin{abstract}

We report 1606 new proper motion systems in the southern sky
(declinations $-$90$\degr$ to $-$47$\degr$) with 0$\farcs$40 yr$^{-1}$
$>$ $\mu$ $\ge$ 0$\farcs$18 yr$^{-1}$.  This effort is a continuation
of the SuperCOSMOS-RECONS (SCR) proper motion search to lower proper
motions than reported in Papers VIII, X, XII, and XV in this series.
Distance estimates are presented for the new systems, assuming that
all stars are on the main sequence.  We find that 31 systems are
within 25 pc, including two systems --- SCR 0838-5855 and SCR
1826-6542 --- we anticipate to be within 10 pc.  These new discoveries
constitute a more than ten-fold increase in new systems found in the
same region of sky searched for systems with $\mu$ $\ge$ 0$\farcs$40
yr$^{-1}$, suggesting a happy hunting ground for new nearby slower
proper motion systems in the region just north (declinations
$-$47$\degr$ to 0$\degr$), much of which has not been rigorously
searched during previous efforts.

\end{abstract}


\section{INTRODUCTION}

In this new edition of the investigation of the solar neighborhood, we
continue the search for nearby stars by focusing our efforts on
systems with proper motions between 0$\farcs$40 yr$^{-1}$ and
0$\farcs$18 yr$^{-1}$.  The likelihood that a slow proper motion
system is nearby is far lower than a fast proper motion system.
Nonetheless, the much larger number of slow proper motion systems
discovered provides an extensive dataset for Galactic structure
analyses, yields many candidates for followup work, and reveals a few
nearby gems that move little relative to the Sun.

The pioneering surveys of
\citep{1971lpms.book.....G,1978LowOB...8...89G} and
\citep{1979lccs.book.....L,1980PMMin..55....1L}\footnote{VizieR Online
Data Catalogue, I/98A \citep{1995yCat.1098....0L}} have provided most
of the proper motion systems catalogued, even as massive computer
searches of digitized photographic plates have become possible.  The
comprehensive New Luyten Two-Tenths catalogue (NLTT) contains 58693
proper motion objects with $\mu$ $\ge$ 0$\farcs$18 yr$^{-1}$.  Since
then, many new high proper motion surveys have been carried out using
new techniques, each of which compliments the work of Giclas and
Luyten.  In chronological order, surveys that sampled the sky south of
declination $-$47$\degr$ (the region relevant to the survey reported
here) include (1) seven papers covering various portions of the
southern sky by Wroblewski and collaborators who used photographic
plates \citep{1994A&AS..105..179W}, (2) UKST survey plates of 40
survey fields by Scholz and collaborators \citep{2000A&A...353..958S,
2002ApJ...565..539S}, (3) a survey of the South Galactic Cap down to
$R_{59F}$ $=$ 19.8 \citep{2001Sci...292..698O}, (4) the
machine-selected catalogue of 11289 objects from SuperCOSMOS R-band
material generated by \citep{2004A&A...421..763P}, (5) our own
SuperCOSMOS-RECONS (SCR) proper motion search of the entire southern
sky
\citep{2004AJ....128..437H,2004AJ....128.2460H,2005AJ....129..413S,2005AJ....130.1658S},
which uses plates with all four emulsions, (6) the southern infrared
proper motion survey (SIPS) \citep{2005A&A...435..363D}, which uses
a combination of Two Micron All Sky Survey (2MASS) data and
SuperCOSMOS I band images, and (7) Lepine's SUPERBLINK survey of a
portion of the southern sky \citep{2005AJ....130.1247L}.  The
Calan-ESO survey \citep{2001ApJS..133..119R}, which identified proper
motions in 14 ESO areas of the southern sky, did not reach as far
south as declination $-$47$\degr$ (but one of the three ESO regions in
\citep{1993AJ....106.2575R} did reach south of $-$47$\degr$, yielding
39 objects).  In addition, Lepine's continuing SUPERBLINK survey will
provide a vast database of proper-motion systems down to 0$\farcs$015
yr$^{-1}$.  He has already published the northern portion of the
survey, which yielded 61,977 objects \citep{2005AJ....129.1483L}.

Our goal since the beginning of the SCR search has been to complete a
comprehensive proper-motion survey of the neglected southern sky.
Adding results from this paper, we have searched the sky from the
south celestial pole to declination $-$47$\degr$ for objects with
10$\farcs$00 yr$^{-1}$ $\ge$ $\mu$ $\ge$ 0$\farcs$18 yr$^{-1}$, where
the lower cutoff has been chosen to match the NLTT.

\section {SEARCH CRITERIA}

This phase of the SCR search uses techniques identical to our earlier
efforts, simply with a lower proper motion cutoff.  Identical
methodology allows us to assess completeness comprehensively and
compare statistics from the various search phases.  The search
techniques are described in detail in Paper VIII in this series
\citep{2004AJ....128..437H}.  Additional phases of the search can be
found in Papers X \citep{2004AJ....128.2460H}, XII
\citep{2005AJ....129..413S}, and XV \citep{2005AJ....130.1658S}.
Briefly, the SCR search utilizes four of the Schmidt survey
photographic plates available in each ESO/SRC survey field, which
provide astrometric and photometric information in the $B_J$, ESO-R,
$R_{59F}$ and $I_{IVN}$ photographic passbands.  Two distinct epochs
in the R band come from the ESO (first epoch; also known as ESO-R) and
UK (second epoch; also known as AAO-R) Schmidt telescope surveys.
Sources must be detected on at least three plates, and are required to
have $R_{59F}$ brighter than magnitude 16.5.

We introduce clear nomenclature here in an effort to sort three categories
of systems revealed during the SCR proper motion search.  MOTION
systems have $\mu$ $\ge$ 1$\farcs$00 yr$^{-1}$ (Paper VIII).  The
1$\farcs$00 yr$^{-1}$ cutoff is convenient, and has been studied in
detail by \citep{2004PhDT........11J}.  SLOWMO systems have
1$\farcs$00 yr$^{-1}$ $>$ $\mu$ $\ge$ 0$\farcs$50 yr$^{-1}$ (Papers
XII and XV).  The 0$\farcs$50 yr$^{-1}$ cutoff has been selected to
match the famous Luyten Half Second (LHS) sample.  MINIMO systems have
0$\farcs$50 yr$^{-1}$ $>$ $\mu$ $\ge$ 0$\farcs$18 yr$^{-1}$, where the
lower cutoff is designed to match that of the NLTT catalogue.  Papers
XII and XV include some MINIMO stars because we pushed to $\mu$ $=$
0$\farcs$40 yr$^{-1}$ to ensure that we picked up any ``LHS
equivalent'' systems with $\mu$ $\sim$ 0$\farcs$50 yr$^{-1}$.  Those
papers also split the southern sky into two portions: Paper XII
covered the sky from declinations $-$90$\degr$ to $-$47$\degr$, while
Paper XV stretched north to the celestial equator.  In this paper, we
match the boundaries of the search in Paper XII and report the
remainder of the MINIMO systems in the range 0$\farcs$40 yr$^{-1}$ $>$
$\mu$ $\ge$ 0$\farcs$18 yr$^{-1}$.

Photographic plates scanned and folded into the SuperCOSMOS database
are 6$\degr$ $\times$ 6$\degr$ with a 0.5$\degr$ overlap of adjacent
fields on each side, providing $\sim$25 square degrees of unique sky
coverage for each field.  Two hundred fields have been included in the
present search, yielding a total coverage of $\sim$5000 square
degrees, corresponding to 12\% of the entire sky.  In this region a
total of 7410 candidate objects were detected, more than four times
the amount of candidates found in Paper XII.  Fields shown as
whitespaces in Figure~\ref{platecoverage} were omitted because of
crowding near the Magellanic Clouds or Galactic Plane, or because
those plate regions had epoch spreads inadequate for reliable proper
motion measurements.  It is worth noting that three more plates have
been omitted in this paper than in Paper XII because longer plate
epoch separations are required to reliably determine slower proper
motions.  The total region omitted amounts to 2.4\% of the entire sky.

After object detection and parameterization (see
\citet{2001A&A....326..1295H}), coordinates are given to each
detection making use of a grid of reference stars with known
coordinates distributed over the plates as described in
\citet{2001MNRAS.326.1315H}.  The default SuperCOSMOS Sky Survey
(SSS) pairing is then used to extract proper motion objects (a full
description can be found in \citep{2001MNRAS.326.1315H}).  The pairing
is set to exclude all images that appear on all four plates having an
astrometric solution that indicates a proper motion less than the cut
off for this search at 0$\farcs$18 yr$^{-1}$ and a goodness-of-fit
parameter of $\chi^2$ $<$ 1.0 \citep{2004AJ....128..437H}.  Each image
either not paired or having inconsistent astrometric solutions due to
incorrect pairing are then processed one at a time using every
possible combination to find a pair out to the upper limit of the SCR
search, 0$\farcs$40 yr$^{-1}$ for this portion of the search.
Proper motions so determined are relative to the mean reference frame
defined for all stars on the plates, because all stars are used to map
out small scale systematic errors in positions due to the photographic
and measurement processes (see for example
\citep{2001MNRAS.326.1315H}).

We continue to use a three-stage sifting method discussed in detail in
Paper XII to remove false detections.  Generally, if a candidate
object survives two color checks and a check of the ellipticity
quality flag, it advances to a check of databases for previous
identifications.  Coordinates of these objects are cross-checked using
both VizieR and SIMBAD to identify previously known objects.  In
VizieR, a ten arcminute radius is used to match SCR detections with
both the NLTT and The Liverpool-Edinburgh High Proper Motion (LEHPM)
catalogues.  If the coordinates of an SCR object are within a few
arcminutes of a catalogue object, and the proper motion and magnitude
match, then the target object is labeled as previously known.  As a
final discovery verification, all potentially new objects are checked
against SIMBAD to determine if they have been previously reported as
proper motion objects.  One class of known objects, GSC objects from
the Hubble Space Telescope (HST) Guide Star Catalogue, is {\it not}
considered previously ``found'', as those stars have merely been
selected as pointing reference points for HST, not as scientifically
noteworthy sources.  Some near matches were found to be common proper
motion (CPM) companions to previously known proper motion objects.
All of these new companions are visually inspected for accuracy and
are discussed in $\S$5.3.

If the object is not found to be previously known, a visual inspection
is done to confirm its veracity.  Visual inspections are carried out
by blinking the $B_J$ and $R_{59F}$ SuperCOSMOS digitally scanned
plate images, which have a sufficient spread in epochs to confirm or refute
derived proper motions.  Thus, all objects reported here have been
confirmed by eye, but some small number of real objects may have been
discarded because visual inspections were not carried out for sources
that did not pass the three checks.

In keeping with previous discovery statistics, we compare the
successful hit rates (real proper motion objects / total candidates
extracted) for the MOTION, SLOWMO, and MINIMO samples in the portion
of the sky covered by all three, from declinations $-$90$\degr$ to
$-$47$\degr$ (not including individual plate regions that were
discarded in any search).  The hit rates are 6.6\% for MOTION systems,
78.6\% for SLOWMO systems, and 78.1\% for MINIMO systems.  These hit rates
take into account new, known, duplicate, and ``garbage'' (not real)
objects.  In Figure~\ref{hitrate}, we show a plot of proper motion
vs.~percent successful hit rate for the entire SCR sample having $\mu$
$\ge$ 0$\farcs$18 yr$^{-1}$ south of declination $-$47$\degr$.  The
SCR search has a sweet spot near $\mu$ $\sim$ 0$\farcs$45 yr$^{-1}$,
where the success rate in picking up proper motion objects is 93\%.
This high success rate deteriorates toward faster and slower proper
motion regimes, where the SCR search is more susceptible to false
detections.  Causes include mistaken object movement from one plate to
another when there are slight shifts between the plates, focus
abnormalities, spurious objects created via plate defects, and bright
star halos.  In Table~\ref{disc-stats}, SCR objects in the various
papers have been categorized into the MOTION, SLOWMO, and MINIMO
samples and discovery statistics are shown for the entire SCR sample.

\section{COMPARISON TO PREVIOUS PROPER-MOTION SURVEYS}

The NLTT catalogue lists 58693 objects with $\mu$ $\ge$ 0$\farcs$18
yr$^{-1}$.  Of these, 2278 objects meet the search criteria of this
paper and have estimated red magnitudes of 16.5 or brighter.  We
recover 1852 of 2278 objects found in the fields we searched,
resulting in an 81\% recovery rate.  Several factors contribute to the
19\% of unrecovered stars.  Our proper motions are accurate to
$\sim$0$\farcs$02 yr$^{-1}$ and our magnitudes to $\sim$0.3 mag.
Proper motions and magnitudes in the NLTT in some cases differ from
ours, which causes us to drop some objects that were included in the
NLTT.  On rare occasions we also miss stars that are lost on recent
plates due to mergers, but were uncorrupted during previous efforts.
Our SCR search has trouble picking up brighter sources because of halo
effects present on the photographic plates.  The brightest NLTT source
we identified in the search is at $R_{59F}$ $\sim$ 5.4 mag while the
brightest NLTT source in this portion of the sky is R $\sim$ 1.6 mag.
Here we compare the results from our SCR survey to date for
declinations $-$90$\degr$ to $-$47$\degr$ to the proper motion surveys
listed in Table~\ref{pm-surveys}\footnote{For samples that were
reported in 1950 coordinates, we have not computed 2000 coordinates or
slid stars because of proper motion over 50 years.  The number of
objects that may have slipped out of the sample by moving north of
$-$47$\degr$ is expected to be small, and a comparable number of
replacement objects likely have slipped south.}.  The numbers listed
reflect only the total number of entries in each work --- we have not
confirmed that all sources by other groups are both real and original.
 
\section {DATA}

Rather than list all of the 1606 new MINIMO systems reported in this
paper explicitly, we give the first five lines in
Table~\ref{scr-tblfull}; the full Table can be found in the electronic
version of this paper.  In Table~\ref{scr-tbl} we highlight the 31
systems for which we estimate distances of less than 25 pc, and 13
additional white dwarf candidates.  As in previous SCR search papers,
in Table~\ref{scr-tblfull} and \ref{scr-tbl} we list SCR names,
coordinates, relative proper motions, plate magnitudes from
SuperCOSMOS, infrared photometry from 2MASS, the $R_{59F}-J$ color, a
distance estimate, and notes.

All coordinates are for epoch and equinox J2000.0 and are computed
using the Two Micron All Sky Survey (2MASS) coordinates and adjusted
to epoch 2000.0 using the SCR proper motions and position angles.
From our search algorithm, the average proper motion errors are
$\sim$0$\farcs$010 yr$^{-1}$.  We show in Figure~\ref{pm-error} that
our proper motion and position angle measurements are consistent with
those in NLTT and Hipparcos by examining 298 MINIMO stars which had
data in both catalogues.\footnote{NLTT proper motion data are
quantized south of DEC $-$45$\degr$ because those proper motions are
reported to hundredths, whereas north of DEC $-$45$\degr$ proper
motions are reported to thousandths.}  The results indicate that the
SCR proper motions and position angles have an average deviation of
$\sim$0$\farcs$025 yr$^{-1}$ and $\sim$6.8$\degr$ compared to NLTT
motions and $\sim$0$\farcs$020 yr$^{-1}$ and $\sim$3.9$\degr$ compared
to Hipparcos.  We also show that the NLTT and Hipparcos proper motions
and position angles show average deviations in proper motions of
$\sim$0$\farcs$019 yr$^{-1}$ and $\sim$5.1$\degr$ when compared
to each other.  Comparison of the three outlying data points in
position angle in the SCR vs.~NLTT and Hipparcos vs.~NLTT plots
indicates that these NLTT measurements are incorrect.  The consistency
between SCR and Hipparcos is particularly encouraging because
Hipparcos only observed stars brighter than $V \sim$ 12, and these are
the stars with the poorest measured proper motions in the SCR survey.

Photographic magnitudes are given in Table~\ref{scr-tblfull} and
\ref{scr-tbl} for three plate emulsions: $B_J$, $R_{59F}$, and
$I_{IVN}$.  Plate magnitude errors are typically less than 0.3 mag for
sources fainter than 15th mag, with errors increasing for brighter
objects because of systematic errors \citep{2001A&A....326..1295H}.
Plate color errors are typically only 0.07 mag.  2MASS $JHK_s$
photometry and a representative color, $R_{59F}-J$, generated using
two of the most reliable photometric measurements, are given.  All
$JHK_{s}$ infrared photometry is extracted from 2MASS via VizieR and
these magnitudes were spot-checked by eye for accuracy.  2MASS
magnitude errors are 0.03 mag or less in most cases.  This holds true
unless $J >$ 15, $H >$ 14.5 or $K_{s}$ $>$ 14, for which the errors
are typically 0.05 mag or greater.

The six-band photometry is used to generate 12 colors that are
utilized to compute distance estimates, as described in
\citet{2004AJ....128..437H}.  All estimates assume that the objects
are single, main sequence stars.  The scatter is 26\%, defined as the
mean of the absolute values of the differences between distances for
stars with trigonometric parallaxes and distances estimated via the
relations.  In cases where objects are too blue for the
relations, no distance is listed.  In a few cases the $B_J$ and/or
$I_{IVN}$ plate magnitudes are not determined (likely causes include
confusing sources and plate defects), making the distance estimate
less reliable.  Thirteen white dwarf and 174 subdwarf candidates have
erroneous distances (listed in brackets) from the suite of relations
because they are not main sequence stars; more accurate estimates are
given in the notes for the white dwarfs.  Overall, we find two systems
within 10 pc, 29 between 10 and 25 pc, and 377 between 25 and 50 pc.

Some of the CPM candidates were not revealed during the SCR search
because of confusion with another source, or because the companion was
fainter than our magnitude cutoff; these were however, noticed during
the blinking process.  In these cases the objects are investigated
using SIMBAD and VizieR to check for previous identifications.  If the
object was not previously known, we use SuperCOSMOS and 2MASS magnitudes
to obtain the six colors used to compute distance estimates, and
proper motion data from SuperCOSMOS to compute proper motions and
position angles.

\section {ANALYSIS}

\subsection {Color-Magnitude Diagram}

Plotted in Figure~\ref{coloram} is a color magnitude diagram comparing
the new SCR objects (open circles) to the known objects (small points)
from this phase of the search.  Data points below $R_{59F} =$ 16.5 are
CPM companions noticed durring the blinking process.  As in Papers XII
and XV, the bulk of the new discoveries are fainter and redder than
the known stars, including three objects, SCR 0838-5855, SCR
1826-6542, and SCR 2241-6119B with $R_{59F}-J >$ 5.5.  Unlike the
sources extracted in our previous efforts, there are more new
discoveries featured in this paper that are brighter and bluer than in
previous discoveries.  In fact, there are 9 new discoveries with
$R_{59F}$ brighter than 10th magnitude that had not been identified
previously.  The point at $R_{59F} =$ 8.14 is SCR 1914-7109, which is
too blue for us to estimate a distance using the plate-2MASS
relations.

As in comparable plots in previous papers, several white dwarf
candidates are clearly separated from the bulk of the sample.  The
subdwarf population is less well-defined, but a population of
subdwarfs can be detected below the concentration of main sequence
stars.

\subsection {Reduced Proper Motion Diagram}

Shown in Figure~\ref{reducedpm} is the reduced proper motion (RPM)
diagram for objects found in this search, similar to the RPM diagrams
shown in Papers XII and XV.  The RPM diagram relies on the statistical
(inverse) relationship between proper motion and distance: objects
that have larger distances generally have smaller proper motions.  As
such, it can be used to separate white dwarfs and (less clearly)
subdwarfs from main-sequence stars.  While this assumption is not
always valid, it is a fairly reliable indicator of luminosity class
for most stars.  As in previous papers, the relation used to determine
H$_{R}$ is

\begin{displaymath}
H_R = R_{59F} + 5 + 5\log\mu.
\end{displaymath}

\noindent A comparison of Figures~\ref{coloram} and \ref{reducedpm}
shows that the RPM diagram is slightly better at separating the three
samples.

There is a clear break separating the white dwarfs from the rest of
the sample.  The arbitrary dashed line between the subdwarfs and white
dwarfs is the same as used in Paper XV, although the proper motions
are smaller.  In this region of the diagram, we find 12 new white
dwarf candidates: SCR 0004-6120B, SCR 0018-6851, SCR 0104-5742B, SCR
0150-7207, SCR 0245-6038, SCR 0355-5611, SCR 0840-7826, SCR 0857-6032,
SCR 1821-5951, SCR 2020-7806, SCR 2032-4948B, and SCR 2354-6023.  One
additional candidate, SCR 0429-5423B, was noticed while blinking to
have the colors consistent with a white dwarf, but lacks the 2MASS
data to plot in the RPM diagram, bringing the total number of white
dwarf candidates for this paper to 13. One object SCR 1800-5112B is a
close double with blended photometry in SuperCOSMOS, which puts it
inside the white dwarf regime of the RPM diagram. The infrared colors
however are not consistent with a white dwarf, and therefore is not
considered a white dwarf candidate.  Of the 13 candidates, three (SCR
0104-5742B, SCR 0150-7207 and SCR 1821-5951) have been
spectroscopically confirmed (results to be presented in a future
publication).  Each is listed in Table~\ref{scr-tbl} with a bracketed
erroneous distance estimate assuming the object is a main sequence
star.  We include a distance estimate based on the assumption of the
source being a main sequence member because large distances can be
used to flag possible white dwarfs. All white dwarfs are estimated to
be beyond 700 pc except SCR 0018-6851, SCR 0150-7207, SCR 0429-5423B,
SCR 0857-6032, and SCR 1821-5951 which have blended SuperCOSMOS
photometry, no 2MASS data, or are too blue for the relations.  In the
notes we give more accurate distance estimates assuming the objects
are single white dwarfs and using the relation of
\citep{2001Sci...292..698O}, with adopted errors of 20\% as given by
the authors.  This brings the total number of white dwarf candidates
to 23 for the SCR survey.

To be consistent with Paper XV, subdwarf candidates are selected to
have $R_{59F}-J >$ 1.0 and $H_{R} >$ 4.0 mag above the dashed line
separating the white dwarfs from the subdwarfs.  As with the white
dwarf cutoff line, the subdwarf cutoff used is arbitrary, yet provides
a fairly accurate distinction between the subdwarfs and main sequence
classes.  Although some contamination of the sample is expected, we
count a total of 174 new subdwarf candidates from this paper.  This
brings the total number of SCR subdwarf candidates to date to 238.

\subsection {New Common Proper Motion Systems}

In this search we find a total of 54 new likely CPM systems, including
52 doubles and two triples (i.e.~56 companions).  All primaries,
companions and their proper motions are recorded in
Table~\ref{CPM-Table}, the separations and position angles are of
companions relative to the primary star.  Distance estimates for
multiple components that agree to within a factor of two are
considered to be consistent, given the errors of the distance
estimating relations.  Twenty-two pairs of objects were discovered via
the automated search and were subsequently noticed to have common
proper motion.  Twenty-five additional new SCR companions not revealed
during the automated search were noticed during the blinking process.
Nine previously known objects were noticed by eye to have common
proper motions to SCR stars discovered during the automated search.

In Figures~\ref{CPM-mu} and \ref{CPM-pa}, we compare the proper
motions and position angles for the 52 pairs having complete sets of
$\mu$ and $\theta$ (because of blending in four systems, SuperCOSMOS
data could not be obtained for the potential companion, and they were
too close to obtain accurate manual measurements).  Values obtained
through the systematic SCR trawl are shown with filled circles, while
open circles are used if a companion was noticed during the blinking
process of the SuperCOSMOS scans.  Data were then retrieved manually
from SuperCOSMOS for the noticed companions.  Usually, such companions
are fainter than our search cutoff of $R_{59F} =$ 16.5.

As is typical with proper motion surveys, the position angles of the
proper motions are better determined than the proper motions
themselves, which are especially volatile at these relatively low
values.  Hence, the position angles are given greater weight when
deciding whether or not two sources really are part of the same
system.  The average scatter for the proper motions and position
angles are $\sim$0$\farcs$020 yr$^{-1}$ and $\sim$ 5.0 degrees,
respectively, which is consistent with comparisons of SCR data to NLTT
and Hipparcos, as discussed previously.

\subsection {Comments on Individual Systems}

{\bf SCR 0838-5855} has $R_{59F} =$ 16.11 and $R_{59F}-J =$ 5.80.
This is likely to be a late M dwarf at an estimated distance of only
8.4 pc.  This is a high priority target on our CTIOPI parallax program
at the CTIO 0.9m \citep{2005AJ....129.1954J,2006AJ....132.2360H}.

{\bf SCR 0927-6239BC} are common proper motion companions to NLTT
21827. The B component has a separation of 200$\farcs$9 at position
angle 92.0$\degr$ from the primary.  The C component is not in 2MASS
and has a separation of 316$\farcs$2 at a position angle of
59.6$\degr$ from the primary. See Table~\ref{CPM-Table}.

{\bf SCR1441-7338} has $R_{59F} =$ 16.14 and $R_{59F}-J =$ 4.95 with a
distance estimate of 19.0 pc.

{\bf SCR 1826-6542} has $R_{59F} =$ 16.43 and $R_{59F}-J =$ 5.86.
This is likely to be a late M dwarf at an estimated distance of only
9.2 pc.  

{\bf SCR 1914-7109} is very bright, with $R_{59F} =$ 8.14 and
$R_{59F}-J =$ $-$0.54.  It is too blue for us to estimate a distance
using the plate-2MASS relations.

{\bf SCR 2057-6358ABC} is a possible triple system with the A and B
component separated by less than 2$\farcs$00, which have blended
SuperCOSMOS photometry, but separated in 2MASS. Both the B and C
components were noticed by eye during the blinking process. The C
component has a separation of 261$\farcs$5 at position angle
121.6$\degr$ from the primary. See Table~\ref{CPM-Table}.

\section {DISCUSSION}

Including all papers pertaining to the SCR proper motion survey, we
have discovered 1967 new systems (2030 objects) with $\mu$ $\ge$ 0$\farcs$18
yr$^{-1}$ between declinations $-$90$\degr$ and 0$\degr$.  Systems
reported in previous papers are represented by filled circles in the
sky map of SCR systems in Figure~\ref{skycoverage}; systems in this
paper are represented by open circles.

The 1606 systems in this paper comprise 82\% of the entire SCR sample
and bring the total number south of $-$47$\degr$ to 1761 systems (1817
objects).  From this paper, there are 13 likely white dwarfs and 174
subdwarf candidates.  Three of the white dwarfs, (but none of the
subdwarfs) have been spectroscopically confirmed, while the rest
remain to be targeted in future spectroscopic efforts to confirm their
luminosity classes.

Discovery statistics for the entire SCR sample, separated by distance
horizons, are given in Table~\ref{diststats}.  In order to be
consistent with previous SCR papers, new common proper motion
candidates that are companions to known objects and probable white
dwarfs (because their distance estimates require a different set of
relations) are not included.  For all proper motion bins except the
last, we have searched the entire southern sky; for the last bin only
$-$47$\degr$ to $-$90$\degr$ has been searched.  In the complete
sample of new SCR systems to date, 7 are estimated to be within 10 pc,
and 67 additional systems lie between 10 and 25 pc.  Of the 1606
systems reported in this paper, two were estimated to be within 10 pc,
with an additional 29 between 10 and 25 pc.  This illustrates that
even in this relatively slow proper motion regime there are nearby
stars hidden in the solar neighborhood.  Should they prove to be
within 10 pc, the two nearest stars reported in this paper would rank
as the 10th (0$\farcs$311 yr$^{-1}$ SCR 1826-6542) and 11th
(0$\farcs$320 yr$^{-1}$ for SCR 0838-5855) slowest systems in the
RECONS sample of 248 systems within 10 pc \citep{2006AJ....132.2360H}.

Although the counts of new 10 pc candidates are small in each proper
motion bin, the largest number of candidates between 10 and 25 pc
is found in the slowest proper motion bin, even though only a portion
of the southern sky has been searched.  We anticipate that continuing
searches for MINIMO systems in the southern sky will be a promising
endeavor for finding additional nearby stars.

\section {ACKNOWLEDGMENTS}

Funding for the SuperCOSMOS Sky Survey was provided by the UK Particle
Physics and Astronomy Research Council.  N.C.H.~would like to thank
colleagues in the Wide Field Astronomy Unit at Edinburgh for their
work in making the SSS possible; particular thanks go to Mike Read,
Sue Tritton, and Harvey MacGillivray.  The RECONS team at Georgia
State University wishes to thank NASA's {\it Space Interferometry
  Mission} and the National Science Foundation (grant AST 05-07711)
for their continued support of our solar neighborhood exploration.
This work has made use of the SIMBAD, VizieR, and Aladin databases
operated at the CDS in Strasbourg, France.  We have also used data
products from the Two Micron All Sky Survey, which is a joint project
of the University of Massachusetts and the Infrared Processing and
Analysis Center, funded by NASA and NSF.


\clearpage


\hoffset-00pt{}
\begin{deluxetable}{lrrr}
\tabletypesize{\tiny}
\tablecaption{Discovery Statistics for Entire SCR Sample\tablenotemark{a}
\label{disc-stats}} 
\tablewidth{0pt} 

\tablehead{\\
\colhead{}&
\colhead{MOTION\tablenotemark{b}}&
\colhead{SLOWMO\tablenotemark{c}}&
\colhead{MINIMO\tablenotemark{d}}}

\startdata

New Discoveries      &    9&  142&  1879 \\
Knowns               &  171& 1159&  5581 \\ 
Duplicates           &   15&   91&   864 \\ 
Garbage              & 1989&  344&  3613 \\ 
\tableline
Total Hits           & 2184& 1736& 11937 \\ 

\enddata
\tablenotetext{a}{Entire SCR Sample, including all previous SCR proper motion papers}
\tablenotetext{b}{MOTION sample includes  $\mu$ $\ge$ 1$\farcs$00 yr$^{-1}$ }
\tablenotetext{c}{SLOWMO sample includes 1$\farcs$00 yr$^{-1}$ $>$ $\mu$ $\ge$ 0$\farcs$50 yr$^{-1}$}
\tablenotetext{d}{MINIMO sample includes 0$\farcs$50 yr$^{-1}$ $>$ $\mu$ $\ge$ 0$\farcs$18 yr$^{-1}$}

\end{deluxetable}

\clearpage


\begin{deluxetable}{lcccccl}
\rotate \tabletypesize{\scriptsize} \tablecaption{Number of NLTT Objects
  Discovered South of Declination $-$47$^\circ$\tablenotemark{a}
\label{pm-surveys}}
\tablewidth{0pt}

\tablehead{\vspace{-8pt} \\
\colhead{Survey}&
\colhead{$\mu$ $\ge$ 1$\farcs$00 yr$^{-1}$}&
\colhead{1$\farcs$00 yr$^{-1}$ $>$ $\mu$ $\ge$ 0$\farcs$50 yr$^{-1}$}&
\colhead{0$\farcs$50 yr$^{-1}$ $>$ $\mu$ $\ge$ 0$\farcs$18 yr$^{-1}$}&
\colhead{Total} &
\colhead{\# of Publications}&
\colhead{References\tablenotemark{b}}}

\startdata
NLTT                               &   56 & 261 & 3529 & 3846 & 1 & 1                \\
SuperCOSMOS-RECONS                 &    5 &  70 & 1742 & 1817 & 5 & 2, 3, 4, 5, 6  \\
Wroblewski and collaborators       &    3 &  28 &  488 &  519 & 4 & 7, 8, 9          \\
SUPERBLINK                         &    1 &  43 &   29 &   73 & 1 & 10               \\
Scholz and collaborators           &    2 &  11 &   55 &   68 & 2 & 11, 12           \\
Ruiz and collaborators             &    0 &   2 &   37 &   39 & 1 & 13               \\
SIPS                               &    7 &  20 &    0 &   27 & 1 & 14               \\
Oppenheimer et al.                 &    1 &   4 &    2 &    7 & 1 & 15               \\
Pokorny et al.                     &mixed &mixed& mixed& mixed& 2 & 16, 17           \\

\enddata 

\tablenotetext{a}{Numbers listed indicate objects reported as ``new''
  in the survey publications.  Pokorny did not differentiate between
  new and previously known objects, so the entry is labled as mixed.}
\tablenotetext{b}{References include (1)
  \citet{1979lccs.book.....L,1980PMMin..55....1L}, (2,3,4,5,6)
  \citet{2004AJ....128..437H,2004AJ....128.2460H,2005AJ....129..413S,2005AJ....130.1658S},
  this paper, (7,8,9)
  \citet{1989A&AS...78..231W,1991A&AS...91..129W,1994A&AS..105..179W},
  (10) \citet{2005AJ....130.1247L}, (11,12)
  \citet{2000A&A...353..958S,2002ApJ...565..539S}, (13)
  \citet{1993AJ....106.2575R}, (14) \citet{2005A&A...435..363D}, (15)
  \citet{2001Sci...292..698O}, (16,17)
  \citet{2003A&A...397..575P,2004A&A...421..763P}}

\end{deluxetable}

\clearpage


\begin{deluxetable}{lcccccrrrrrrrl}
\rotate \tabletypesize{\scriptsize} \tablecaption{Characteristics of
  New SCR Systems with 0$\farcs$40 yr$^{-1}$ $>$ $\mu$ $\ge$
  0$\farcs$18 yr$^{-1}$ from $-$90$\degr$ $<$ $\delta$ $\le$
  $-$47$\degr$
\label{scr-tblfull}}
\tablewidth{0pt}

\tablehead{\vspace{-8pt} \\
\colhead{Name}&
\colhead{RA}&
\colhead{DEC}&
\colhead{$\mu$}&
\colhead{$\theta$}&
\colhead{$B_J$}&
\colhead{$R_{59F}$}&
\colhead{$I_{IVN}$}&
\colhead{$J$}&
\colhead{$H$}&
\colhead{$K_s$}&
\colhead{$R_{59F}$ $-$ $J$}&
\colhead{Est Dist}&
\colhead{Notes}\\

\colhead{}&
\colhead{(J2000)}&
\colhead{(J2000)}&
\colhead{($\arcsec$)}&
\colhead{($\degr$)}&
\colhead{}&
\colhead{}&
\colhead{}&
\colhead{}&
\colhead{}&
\colhead{}&
\colhead{}&
\colhead{(pc)}&
\colhead{}}

\startdata
\vspace{0pt} \\

SCR 0001-7015 & 00 01 56.59 & -70 15 08.4 & 0.202 & 141.1 & 18.22 & 16.20 & 14.64 & 13.36 & 12.79 & 12.66 &  2.84 &  142.5 &                                                                                         \\                                      
SCR 0003-6021 & 00 03 59.01 & -60 21 58.1 & 0.249 & 145.1 & 17.84 & 15.68 & 12.92 & 11.68 & 11.08 & 10.75 &  4.00 &   35.3 &                                                                                         \\
SCR 0004-5740 & 00 04 26.79 & -57 40 02.7 & 0.181 & 091.2 & 16.00 & 13.87 & 12.36 & 11.47 & 10.87 & 10.65 &  2.41 &   66.0 &                                                                                         \\
SCR 0004-6120A& 00 04 56.20 & -61 20 58.0 & 0.180 & 129.0 & 13.63 & 11.28 & 9.94  & 9.66  & 9.08  & 8.81  &  1.62 &   35.8 &  \tablenotemark{a}                                                                      \\
SCR 0004-6120B& 00 04 45.41 & -61 23 40.0 & 0.171 & 127.6 & 16.87 & 16.76 & 16.54 & 16.44 & 15.94 & 16.48 &  0.33 &[1216.2]&  \tablenotemark{a}, \tablenotemark{b}, \tablenotemark{c}                                \\

\enddata
                             
\tablenotetext{a}{Common proper motion companion; see Table~\ref{CPM-Table}}
\tablenotetext{b}{White dwarf candidate picked from RPM diagram, plate distance [in bracket] is incorrect; see Table~\ref{scr-tbl}}
\tablenotetext{c}{Not detected during automated search but noticed by eye during the blinking process}
\tablenotetext{d}{Subdwarf candidate picked from RPM diagram; plate distance [in bracket] is incorrect}
\tablenotetext{e}{Source not in 2MASS}

\end{deluxetable}

\clearpage


\begin{deluxetable}{lccccccccccccl}
\rotate \tabletypesize{\scriptsize} \tablecaption{Characteristics of
  New SCR Systems Estimated to be Nearer than 25 Parsecs and New White
  Dwarf Candidates with 0$\farcs$40 yr$^{-1}$ $>$ $\mu$ $\ge$
  0$\farcs$18 yr$^{-1}$ from $-$90$^\circ$ $<$ $\delta$ $\le$
  $-$47$^\circ$
\label{scr-tbl}}
\tablewidth{0pt}

\tablehead{\vspace{-8pt} \\
\colhead{Name}&
\colhead{RA}&
\colhead{DEC}&
\colhead{$\mu$}&
\colhead{$\theta$}&
\colhead{$B_J$}&
\colhead{$R_{59F}$}&
\colhead{$I_{IVN}$}&
\colhead{$J$}&
\colhead{$H$}&
\colhead{$K_s$}&
\colhead{$R_{59F}$ $-$ $J$}&
\colhead{Est Dist}&
\colhead{Notes}\\

\colhead{}&
\colhead{(J2000)}&
\colhead{(J2000)}&
\colhead{($\arcsec$)}&
\colhead{($\degr$)}&
\colhead{}&
\colhead{}&
\colhead{}&
\colhead{}&
\colhead{}&
\colhead{}&
\colhead{}&
\colhead{(pc)}&
\colhead{}}

\startdata
\vspace{-11pt} \\
\multicolumn{14}{c}{SCR Red Dwarf candidates $<$ 25pc}\\
\tableline\\
SCR 0100-7904 & 01 00 56.08 & -79 04 25.2 & 0.379 & 215.3 & 13.69 & 11.69 &  9.98 &  8.79 &  8.16 &  7.88 & 2.89 &   14.9 &                                                                               \\
SCR 0135-6127 & 01 35 53.66 & -61 27 11.1 & 0.255 & 256.8 & 15.61 & 13.67 & 11.80 & 10.05 &  9.53 &  9.24 & 3.61 &   20.8 &                                                                               \\
SCR 0138-5353 & 01 38 20.51 & -53 53 26.1 & 0.297 & 071.0 & 15.70 & 13.69 & 11.73 & 10.27 &  9.69 &  9.42 & 3.42 &   24.2 &                                                                               \\
SCR 0211-6108 & 02 11 35.42 & -61 08 53.8 & 0.234 & 060.8 & 11.68 &  9.73 &  9.23 &  8.67 &  8.15 &  8.07 & 1.06 &   22.9 &                                                                               \\
SCR 0232-8458 & 02 32 50.12 & -84 58 09.5 & 0.220 & 141.9 & 12.31 & 10.17 &  9.87 &  8.99 &  8.34 &  8.18 & 1.17 &   24.9 &                                                                               \\
SCR 0246-7024 & 02 46 02.25 & -70 24 06.3 & 0.259 & 113.2 & 15.71 & 13.32 & 10.71 &  9.83 &  9.32 &  9.01 & 3.49 &   20.0 &                                                                               \\
SCR 0527-7231 & 05 27 06.99 & -72 31 20.0 & 0.368 & 018.3 & 16.01 & 13.96 & 11.76 & 10.33 &  9.76 &  9.47 & 3.63 &   22.6 &                                                                               \\
SCR 0635-6722 & 06 35 48.81 & -67 22 58.5 & 0.383 & 340.0 & 12.21 &  9.83 &  8.66 &  8.54 &  7.95 &  7.69 & 1.29 &   22.7 &                                                                               \\
SCR 0838-5855 & 08 38 02.24 & -58 55 58.7 & 0.320 & 188.9 & 18.44 & 16.11 & 12.43 & 10.30 &  9.70 &  9.26 & 5.80 &    8.4 &                                                                               \\
SCR 1147-5504 & 11 47 52.49 & -55 04 11.9 & 0.192 & 011.3 & 14.95 & 12.23 & 10.25 &  9.67 &  9.08 &  8.81 & 2.56 &   24.1 &                                                                               \\
SCR 1217-7810 & 12 17 26.93 & -78 10 45.9 & 0.212 & 056.6 & 17.54 & 15.69 & 13.14 & 11.19 & 10.64 & 10.35 & 4.49 &   24.5 &                                                                               \\
SCR 1220-8302 & 12 20 03.71 & -83 02 29.2 & 0.243 & 244.2 & 17.03 & 14.93 & 12.79 & 10.97 & 10.39 & 10.07 & 3.96 &   25.0 &                                                                               \\
SCR 1224-5339 & 12 24 24.44 & -53 39 08.8 & 0.189 & 251.9 & 16.93 & 14.77 & 12.30 & 10.50 &  9.93 &  9.64 & 4.26 &   18.1 &                                                                               \\
SCR 1347-7610 & 13 47 56.80 & -76 10 20.0 & 0.194 & 089.7 & 12.39 & 10.26 &  8.88 &  8.62 &  8.01 &  7.77 & 1.64 &   22.5 &                                                                               \\
SCR 1420-7516 & 14 20 36.84 & -75 16  5.9 & 0.195 & 243.7 & 14.22 & 12.68 & 10.44 &  9.44 &  8.90 &  8.63 & 3.23 &   21.3 &                                                                               \\
SCR 1441-7338 & 14 41 14.42 & -73 38 41.4 & 0.207 & 029.0 & 18.27 & 16.14 & 13.04 & 11.19 & 10.60 & 10.26 & 4.95 &   19.0 &                                                                               \\
SCR 1448-5735 & 14 48 39.82 & -57 35 17.7 & 0.202 & 188.8 & 12.47 & 10.59 & 12.46 &  9.14 &  8.55 &  8.43 & 1.45 &   18.2 &                                                                               \\
SCR 1456-7239 & 14 56 02.29 & -72 39 41.4 & 0.207 & 225.0 & 16.50 & 14.22 & 11.99 & 10.62 & 10.05 &  9.74 & 3.60 &   24.9 &                                                                               \\
SCR 1738-5942 & 17 38 41.02 & -59 42 24.4 & 0.280 & 148.2 & 16.57 & 14.20 & 11.92 & 10.38 &  9.83 &  9.58 & 3.82 &   20.8 &                                                                               \\
SCR 1820-6225 & 18 20 49.35 & -62 25 52.7 & 0.190 & 164.8 & 13.35 & 11.17 &  8.43 &  9.13 &  8.48 &  8.29 & 2.04 &   22.4 &                                                                               \\
SCR 1826-6542 & 18 26 46.83 & -65 42 39.9 & 0.311 & 178.9 & 18.68 & 16.43 & 12.91 & 10.56 &  9.96 &  9.54 & 5.86 &    9.2 &                                                                               \\
SCR 1853-7537 & 18 53 26.61 & -75 37 39.8 & 0.304 & 168.7 & 12.20 &  9.85 &  9.09 &  8.34 &  7.73 &  7.49 & 1.51 &   20.0 &                                                                               \\
SCR 1856-4704 & 18 56 38.40 & -47 04 58.3 & 0.252 & 131.3 & 16.29 & 13.92 & 11.64 & 10.29 &  9.74 &  9.44 & 3.63 &   21.4 &                                                                               \\
SCR 1926-8216 & 19 26 48.64 & -82 16 47.6 & 0.195 & 172.5 & 12.22 & 10.00 & 10.20 &  9.04 &  8.42 &  8.30 & 0.96 &   17.3 &                                                                               \\
SCR 1932-5005 & 19 32 48.64 & -50 05 38.9 & 0.257 & 157.5 & 16.86 & 14.50 & 11.99 & 10.75 & 10.11 &  9.84 & 3.75 &   24.4 &                                                                               \\
SCR 1959-6236 & 19 59 33.55 & -62 36 13.4 & 0.189 & 288.7 & 17.47 & 15.35 & 12.67 & 11.06 & 10.49 & 10.22 & 4.29 &   24.3 &                                                                               \\
SCR 2016-7531 & 20 16 11.25 & -75 31 04.5 & 0.253 & 081.3 & 17.06 & 14.74 & 12.24 & 10.46 &  9.86 &  9.50 & 4.28 &   16.2 &                                                                               \\
SCR 2042-5737 & 20 42 46.44 & -57 37 15.3 & 0.264 & 142.6 & 15.06 & 13.22 & 11.55 &  9.97 &  9.52 &  9.03 & 3.25 &   22.7 &                                                                               \\
SCR 2230-5244 & 22 30 27.95 & -52 44 29.1 & 0.369 & 125.7 & 19.01 & 16.34 &\nodata& 11.84 & 11.23 & 10.91 & 4.49 &   24.6 &                                                                               \\
SCR 2241-6119A& 22 41 44.36 & -61 19 31.2 & 0.184 & 124.0 & 15.64 & 13.65 & 11.70 & 10.20 &  9.61 &  9.35 & 3.44 &   23.2 & \tablenotemark{a}                                                                                     \\
SCR 2335-6433A& 23 35 18.43 & -64 33 42.4 & 0.196 & 103.1 & 11.79 &  9.97 &  9.01 &  8.63 &  8.02 &  7.86 & 1.33 &   24.5 & \tablenotemark{a}                                                                             \\
\tableline\\
\vspace{-15pt}\\
\multicolumn{14}{c}{SCR WD Candidates}\\
\vspace{-8pt}\\
\tableline\\                                                                                                                   
SCR 0004-6120B& 00 04 45.41 & -61 23 40.0 & 0.171 & 127.6 & 16.86 & 16.76 & 16.53 & 16.43 & 15.93 & 16.47 & 0.32 & [1216.2]& \tablenotemark{a}, \tablenotemark{b}, \tablenotemark{c}, 59.3 $\pm$ 11.9pc   \\
SCR 0018-6851 & 00 18 08.56 & -68 51 19.4 & 0.220 & 091.6 & 16.55 & 16.46 & 16.48 & 16.62 & 16.13 & 17.18 & -0.16& \nodata & \tablenotemark{c}, \tablenotemark{e}, 52.2 $\pm$ 10.4pc                      \\
SCR 0104-5742B& 01 04 12.14 & -57 42 48.6 & 0.239 & 091.1 & 16.19 & 15.89 & 15.78 & 15.67 & 15.56 & 15.75 & 0.22 &  [872.1]& \tablenotemark{a}, \tablenotemark{c}, \tablenotemark{f}, 34.5 $\pm$ 6.9pc   \\
SCR 0150-7207 & 01 50 38.49 & -72 07 16.8 & 0.334 & 223.9 & 16.18 & 15.71 & 15.24 & 15.65 & 15.64 & 15.42 & 0.06 & \nodata & \tablenotemark{c}, \tablenotemark{e}, \tablenotemark{f}, 28.0 $\pm$ 5.6pc  \\
SCR 0245-6038 & 02 45 27.77 & -60 38 58.2 & 0.196 & 049.6 & 17.15 & 16.36 & 16.07 & 15.83 & 15.47 & 15.66 & 0.52 &  [821.9]& \tablenotemark{c}, 29.9 $\pm$ 6.0pc                                        \\
SCR 0355-5611 & 03 55 31.89 & -56 11 28.2 & 0.279 & 029.1 & 17.36 & 16.46 & 16.11 & 16.05 & 15.53 & 15.44 & 0.41 &  [755.9]& \tablenotemark{c}, 28.9 $\pm$ 5.8pc                                        \\
SCR 0429-5423B& 04 29 05.93 & -54 23 03.6 & 0.170 &  39.7 & 17.91 & 17.08 & 16.97 &\nodata&\nodata&\nodata&\nodata& \nodata& \tablenotemark{a}, \tablenotemark{b}, \tablenotemark{c}, \tablenotemark{d}, 40.5 $\pm$ 8.1pc \\
SCR 0840-7826 & 08 40 29.00 & -78 26 46.0 & 0.399 & 010.3 & 16.06 & 15.82 & 15.77 & 15.62 & 15.57 & 15.47 & 0.20 &  [763.9]& \tablenotemark{c}, 34.8 $\pm$ 7.0pc                                        \\
SCR 0857-6032 & 08 57 08.21 & -60 32 45.4 & 0.217 & 333.3 & 15.20 & 15.37 & 15.45 & 15.94 & 16.20 & 15.78 & -0.56&  \nodata& \tablenotemark{c}, \tablenotemark{e}, 38.2 $\pm$ 7.6pc                       \\
SCR 1821-5951 & 18 21 59.54 & -59 51 48.5 & 0.365 & 194.9 & 17.49 & 16.31 & 15.72 & 15.20 & 15.00 & 14.90 & 1.11 &  [588.9]& \tablenotemark{c}, \tablenotemark{f}, 22.2 $\pm$ 4.4pc                     \\
SCR 2020-7806 & 20 20 52.98 & -78 06 18.7 & 0.276 & 209.2 & 16.03 & 16.09 & 16.11 & 15.92 & 15.59 & 15.68 & 0.16 &  [834.9]& \tablenotemark{c}, 49.1 $\pm$ 9.8pc                                        \\
SCR 2032-4948B& 20 32 41.74 & -49 48 57.2 & 0.270 & 182.4 & 17.15 & 16.77 & 16.73 & 16.62 & 15.87 & 15.88 & 0.14 & [900.1] & \tablenotemark{a}, \tablenotemark{b}, \tablenotemark{c}, 48.7 $\pm$ 9.7pc                      \\
SCR 2354-6023 & 23 54 50.63 & -60 23 16.0 & 0.230 & 098.6 & 16.31 & 16.06 & 15.93 & 15.87 & 15.77 & 16.31 & 0.18 & [1127.8]& \tablenotemark{c}, 38.6 $\pm$ 7.7pc                                        \\

\enddata                             
\tablenotetext{a}{Common proper motion; see Table~\ref{CPM-Table}}
\tablenotetext{b}{Not detected during automated search but noticed by eye during the blinking process}
\tablenotetext{c}{WD candidate with unreliable distance estimate [in brackets]; more reliable WD distance estimate in notes}
\tablenotetext{d}{Not in 2MASS}
\tablenotetext{e}{All colors are too blue or too red for distance relations}
\tablenotetext{f}{WD confirmed via spectroscopy; results to be published in a future paper by co-author Subasavage}
\end{deluxetable}

\clearpage

\begin{deluxetable}{lccclcccccl}
\rotate 
\tabletypesize{\scriptsize} 
\tablecaption{Common Proper Motion Systems \label{CPM-Table}}
\tablewidth{0pt}
\tablehead{
\colhead{Primary}&
\colhead{$\mu$}&
\colhead{$\theta$}&
\colhead{Distance}&
\colhead{Secondary/Tertiary}&
\colhead{$\mu$}&
\colhead{$\theta$}&
\colhead{Distance}&
\colhead{Separation}&
\colhead{$\theta$}&
\colhead{notes}\\
\colhead{}&
\colhead{($\arcsec$)}&
\colhead{($\degr$)}&
\colhead{(pc)}&
\colhead{} &
\colhead{($\arcsec$)}&
\colhead{($\degr$)}&
\colhead{(pc)}&
\colhead{($\arcsec$)}&
\colhead{($\degr$)}&
\colhead{}}
\startdata
HD 120056      & 0.164 & 230.1 &\nodata& SCR 1348-5536B &  0.204 & 112.1 &  236.9 &  46.9 & 310.9 & \tablenotemark{a}, \tablenotemark{b}, Hipparcos distance at 100.70pc       \\
HD 158866      & 0.235 & 188.4 & (13.1)& SCR 1746-8211B &  0.228 & 184.9 &   14.5 &  76.5 & 290.8 & \tablenotemark{c}, Hipparcos distance at 30.61pc                           \\
NLTT 01733     & 0.250 & 115.3 &  36.1 & SCR 0032-5528B &  0.221 & 115.2 &   70.3 & 349.6 &  71.4 & \tablenotemark{b}                                                          \\
NLTT 03566     & 0.257 &  87.5 &  27.9 & SCR 0104-5742B &  0.239 &  91.1 & [872.1]& 110.6 & 206.6 & \tablenotemark{d}, WD candidate at 34.5 $\pm$ 6.9pc                     \\
NLTT 13852     & 0.233 &  46.3 &\nodata& SCR 0436-8233B &  0.233 &  51.1 &   48.6 & 108.9 & 338.7 & \tablenotemark{a}, Hipparcos distance at 46.46pc                                              \\
NLTT 15903     & 0.223 &  17.9 &  36.4 & SCR 0551-8116B &  0.233 &  19.5 &   56.5 &  99.6 & 266.1 &                                                                            \\
NLTT 20147     & 0.222 & 342.8 &(40.4) & SCR 0843-5007B &  0.225 & 338.8 &   74.1 &  24.0 &  64.3 & \tablenotemark{b}, \tablenotemark{c}, Hipparcos distance at 93.63pc        \\
NLTT 21827     & 0.231 & 312.1 &  53.9 & SCR 0927-6239B &  0.204 & 312.2 &   41.6 & 200.9 &  92.0 &                                                                            \\
               &       &       &       & SCR 0927-6239C &  0.197 & 309.0 & \nodata& 316.2 &  59.6 & \tablenotemark{b}, \tablenotemark{e}                                       \\
NLTT 26256     & 0.194 & 283.9 &(12.5) & SCR 1104-7856B &  0.203 & 283.1 &  101.1 & 260.4 & 167.7 & \tablenotemark{b}, \tablenotemark{c}, Hipparcos distance at 47.46pc        \\
NLTT 36394     & 0.238 & 238.9 & (36.9)& SCR 1411-7525B &  0.250 & 242.1 &  111.5 &  21.8 & 275.5 & \tablenotemark{c}                                                          \\
NLTT 45592     & 0.174 & 178.9 & (23.3)& SCR 1755-7241B &  0.199 & 175.2 &   60.3 & 102.2 & 273.7 & \tablenotemark{b}, \tablenotemark{c}, Hipparcos distance of 77.94pc        \\
NLTT 49033     & 0.270 & 159.1 &  28.8 & SCR 2019-4701B &  0.265 & 158.1 &   52.5 &  67.9 &  86.9 & \tablenotemark{b}                                                          \\
NLTT 49299     & 0.295 & 100.1 & (22.9)& SCR 2029-5757B &  0.314 & 107.8 &   79.0 & 111.7 & 102.8 & \tablenotemark{c},  Hipparcos distance at 56.31pc                          \\
PPM 365916     & 0.259 & 135.1 & (33.0)& SCR 2237-6238B &  0.191 & 127.4 &   69.6 & 118.8 & 107.6 & \tablenotemark{c}                                                          \\
SCR 0004-6120A & 0.180 & 129.0 &  35.8 & SCR 0004-6120B &  0.171 & 127.6 &[1216.2]& 179.6 & 205.5 & \tablenotemark{b}, \tablenotemark{d}, WD candidate at 59.3 $\pm$ 11.9pc   \\
SCR 0017-5036A & 0.204 & 112.1 & (83.6)& NLTT 00918     &\nodata &\nodata& \nodata&\nodata&\nodata& \tablenotemark{a}, \tablenotemark{b}, \tablenotemark{c}                    \\
SCR 0025-5254A & 0.196 & 101.2 &  44.4 & SCR 0025-5254B &  0.110 & 107.1 &   80.0 & 286.8 & 330.7 & \tablenotemark{b}                                                          \\
SCR 0055-5529A & 0.234 &  53.6 &  88.9 & SCR 0055-5529B &  0.212 &  52.6 & \nodata&   5.8 & 359.3 & \tablenotemark{a}, \tablenotemark{b}                                       \\
SCR 0156-6402A & 0.197 &  85.6 & 140.4 & SCR 0156-6402B &  0.217 &  86.2 &   82.1 & 280.5 & 261.8 & \tablenotemark{b}                                                          \\
SCR 0156-6702A & 0.183 &  57.4 &  29.7 & SCR 0156-6702B &  0.177 &  47.4 &   45.2 &  19.3 &  79.8 & \tablenotemark{b}                                                          \\
SCR 0205-6122A & 0.184 & 192.8 & [74.6]& SCR 0205-6122B &\nodata &\nodata& \nodata&\nodata&\nodata& \tablenotemark{a}, \tablenotemark{b}                                       \\
SCR 0206-6609A & 0.181 &  80.6 & 101.4 & SCR 0206-6609B &  0.220 &  75.9 &  124.5 & 107.4 &  92.6 & \tablenotemark{b}                                                          \\
SCR 0246-4935A & 0.247 &  72.3 &  40.0 & SCR 0246-4935B &  0.243 &  58.0 &  (31.4)&   8.2 &  145.3& \tablenotemark{a}, \tablenotemark{b}, \tablenotemark{c}                    \\
SCR 0353-6844A & 0.232 &  38.0 &  53.2 & SCR 0353-6844B &  0.238 &  38.1 &   59.3 &  13.5 & 226.4 &                                                                            \\
SCR 0429-5423A & 0.188 &  49.4 &  54.4 & SCR 0429-5423B &  0.170 &  39.7 &\nodata &138.0 & 115.4 & \tablenotemark{b}, \tablenotemark{d}, WD candidate at 40.5 $\pm$ 8.1pc     \\
SCR 0454-8045A & 0.195 &  14.2 &  43.0 & SCR 0454-8045B &  0.191 &  13.5 &   56.5 &  14.4 & 189.1 &                                                                            \\
SCR 0611-7302A & 0.188 & 208.1 &  65.7 & SCR 0611-7302B &  0.190 & 211.4 &   69.4 & 101.9 &  46.1 &                                                                            \\
SCR 0745-4814A & 0.200 & 174.9 &  31.4 & SCR 0745-4814B &  0.201 & 176.0 &   25.7 &  43.7 & 220.5 & \tablenotemark{b}                                                          \\
SCR 0757-7444A & 0.197 &  50.9 & (99.4)& SCR 0757-7444B &  0.180 &  48.0 &  158.5 &  10.4 & 311.4 & \tablenotemark{b}, \tablenotemark{c}                                       \\
SCR 0853-7705A & 0.211 & 306.7 &  52.6 & SCR 0853-7705B &  0.225 & 293.2 &   76.2 & 224.1 &  99.5 &                                                                            \\
SCR 0921-7523A & 0.291 & 321.1 &  62.2 & SCR 0921-7523B &  0.286 & 316.3 &   80.4 &  14.0 & 309.5 &                                                                            \\
SCR 1117-7226A & 0.188 & 275.8 &  72.2 & NLTT 26927     &  0.204 & 273.7 &   52.0 &  28.8 &  30.5 &                                                                            \\
SCR 1229-5738A & 0.197 & 256.6 &  92.7 & SCR 1229-5738B &  0.189 & 262.7 &  109.8 &  10.8 & 215.0 & \tablenotemark{b}                                                          \\
SCR 1246-5328A & 0.188 &  29.0 &  45.3 & SCR 1246-5328B &  0.183 &  25.3 &   82.1 &  19.5 & 298.8 &                                                                            \\
SCR 1800-5112A & 0.289 & 226.5 &\nodata& SCR 1800-5112B &  0.317 & 220.3 &  178.3 &   4.6 & 123.7 & \tablenotemark{a}, \tablenotemark{b}                                       \\
SCR 1804-5541A & 0.222 & 194.4 & (47.2)& SCR 1804-5541B &  0.198 & 201.7 &  163.4 &  18.9 & 302.6 & \tablenotemark{b}, \tablenotemark{c}                                       \\
SCR 1809-6154A & 0.218 & 253.2 & 159.3 & SCR 1809-6154B &  0.201 & 259.8 &  163.9 &  19.3 & 269.7 &                                                                            \\
SCR 1816-6615A & 0.209 & 207.1 &[448.4]& SCR 1816-6615B &\nodata &\nodata& \nodata&\nodata&\nodata& \tablenotemark{a}, \tablenotemark{b}                                       \\
SCR 1902-7550A & 0.303 & 168.8 &  75.9 & P 1248-2       &  0.286 & 168.5 &  111.7 &  20.3 & 129.4 & \tablenotemark{b}                                                          \\
SCR 1912-4910A & 0.229 & 187.7 &  35.8 & SCR 1912-4910B &  0.254 & 185.5 &   42.3 &  20.9 & 262.8 &                                                                            \\
SCR 1917-4915A & 0.216 & 200.3 &  48.2 & P 3221-2       &  0.212 & 200.8 &   81.8 &  44.0 & 131.6 & \tablenotemark{b}                                                          \\
SCR 1953-5037A & 0.181 & 144.3 &  31.6 & SCR 1953-5037B &  0.171 & 142.9 &   45.0 &  82.5 & 188.8 & \tablenotemark{b}                                                          \\
SCR 1958-8000A & 0.183 & 149.1 & (61.3)& SCR 1958-8000B &  0.192 & 145.8 & (133.4)&   6.7 & 258.7 & \tablenotemark{b}, \tablenotemark{f}, without I A at 81pc and B at 128pc   \\
SCR 2032-4948A & 0.265 & 179.0 &  75.3 & SCR 2032-4948B &  0.270 & 182.4 &[900.1] &   8.8 & 213.1 & \tablenotemark{b}, \tablenotemark{d}, WD candidate at 48.7 $\pm$ 9.7pc   \\
SCR 2036-6454A & 0.197 & 146.0 &  72.7 & SCR 2036-6454B &  0.190 & 153.3 &   88.2 &  35.3 &  72.5 &                                                                            \\
SCR 2057-6358A & 0.187 & 139.9 &[102.0]& SCR 2057-6358B &\nodata &\nodata& \nodata&\nodata&\nodata& \tablenotemark{a}, \tablenotemark{b}                                       \\
               &       &       &       & SCR 2057-6358C &  0.130 & 138.7 &   46.7 & 261.5 & 121.6 & \tablenotemark{b}                                                          \\
SCR 2100-5804A & 0.186 & 117.6 &  80.5 & SCR 2100-5804B &  0.181 & 116.1 &  147.0 &  16.4 & 300.7 & \tablenotemark{b}                                                          \\
SCR 2112-5428A & 0.209 &  95.3 &  35.6 & SCR 2112-5428B &  0.152 & 117.0 &  13.5  &  15.3 &  60.6 & \tablenotemark{b}                                                          \\
SCR 2218-5310A & 0.190 &  95.3 & (61.1)& SCR 2218-5310B &  0.191 &  95.9 &   96.6 &  35.2 & 142.1 & \tablenotemark{c}                                                          \\
SCR 2241-6119A & 0.184 & 124.0 &  23.2 & SCR 2241-6119B &  0.162 & 117.6 &   27.8 &   9.6 & 211.2 & \tablenotemark{b}                                                          \\
SCR 2255-5120A & 0.335 & 227.1 &  69.7 & SCR 2255-5120B &  0.263 & 228.9 &  112.6 &  17.4 & 359.3 & \tablenotemark{b}                                                          \\
SCR 2333-7923A & 0.394 & 101.4 &[183.1]& SCR 2333-7923B &  0.388 & 100.0 & [216.2]&  11.2 & 123.0 & \tablenotemark{g}                                                          \\
SCR 2335-6433A & 0.196 & 103.1 & (24.5)& SCR 2335-6433B &  0.196 &  99.1 &   57.5 &  22.4 &  25.9 & \tablenotemark{h}                                                          \\
SCR 2353-8204A & 0.216 &  66.9 & (26.8)& SCR 2353-8204B &  0.209 &  68.4 &   54.2 &  34.5 & 181.9 & \tablenotemark{h}                                                          \\

\enddata
\tablenotetext{a}{Distance estimate unreliable when listed [in brackets]; if distance not given BRI photometry not available}
\tablenotetext{b}{Companion not detected during automated search but noticed to be a common proper motion companion during visual inspection}
\tablenotetext{c}{Fewer than 6 relations, therefore distance estimate unreliable (in parenthesis); Hipparcos distance given in notes when available}
\tablenotetext{d}{White dwarf candidate with unreliable distance when listed [in brackets]}
\tablenotetext{e}{No 2MASS photometry available for companion}
\tablenotetext{f}{I photometry suspect, therefore distance unreliable (in parenthesis)} 
\tablenotetext{g}{Subdwarf candidate with unreliable distance listed [in brackets]}
\tablenotetext{h}{Distance estimate unreliable (in parenthesis), because primary is likely a double}

\end{deluxetable}

\clearpage


\hoffset-00pt{}
\begin{deluxetable}{cccc}
\tabletypesize{\tiny}
\tablecaption{Distance Estimate Statistics for New SCR Systems.\tablenotemark{a}
\label{diststats}}
\tablewidth{0pt}

\tablehead{\vspace{-8pt} \\
\colhead{Proper motion}&
\colhead{d $\leq$ 10 pc}&
\colhead{10 pc $<$ d $\leq$ 25 pc}&
\colhead{d $>$ 25 pc}}

\startdata
$\mu$ $\geq$ 1$\farcs$00 yr$^{-1}$                         & 2 $+$ 0 $+$ 0 &  0 $+$  0 $+$  0 &   2 $+$   4 $+$    0 \\
1$\farcs$00 yr$^{-1}$ $>$ $\mu$ $\geq$ 0$\farcs$80 yr$^{-1}$& 0 $+$ 0 $+$ 0 &  3 $+$  0 $+$  0 &   2 $+$   1 $+$    0 \\
0$\farcs$80 yr$^{-1}$ $>$ $\mu$ $\geq$ 0$\farcs$60 yr$^{-1}$& 0 $+$ 1 $+$ 0 &  4 $+$  7 $+$  0 &  25 $+$  23 $+$    0 \\
0$\farcs$60 yr$^{-1}$ $>$ $\mu$ $\geq$ 0$\farcs$40 yr$^{-1}$& 1 $+$ 1 $+$ 0 &  8 $+$ 16 $+$  0 &  95 $+$  93 $+$    0 \\
0$\farcs$40 yr$^{-1}$ $>$ $\mu$ $\geq$ 0$\farcs$18 yr$^{-1}$& 0 $+$ 0 $+$ 2 &  0 $+$  0 $+$ 29 &   0 $+$   0 $+$ 1542 \\
\tableline
\vspace{-3pt}
Total                                                     & 3 $+$ 2 $+$ 2 & 15 $+$ 23 $+$ 29 & 124 $+$ 121 $+$ 1542 \\
\enddata

\tablenotetext{a}{Entire SCR sample excluding white dwarf candidates
  and new common proper motion companions to known objects noticed by
  eye; the first number from Paper XII, second number from Paper XV,
  third number from this paper}

\end{deluxetable}

\clearpage


\figcaption[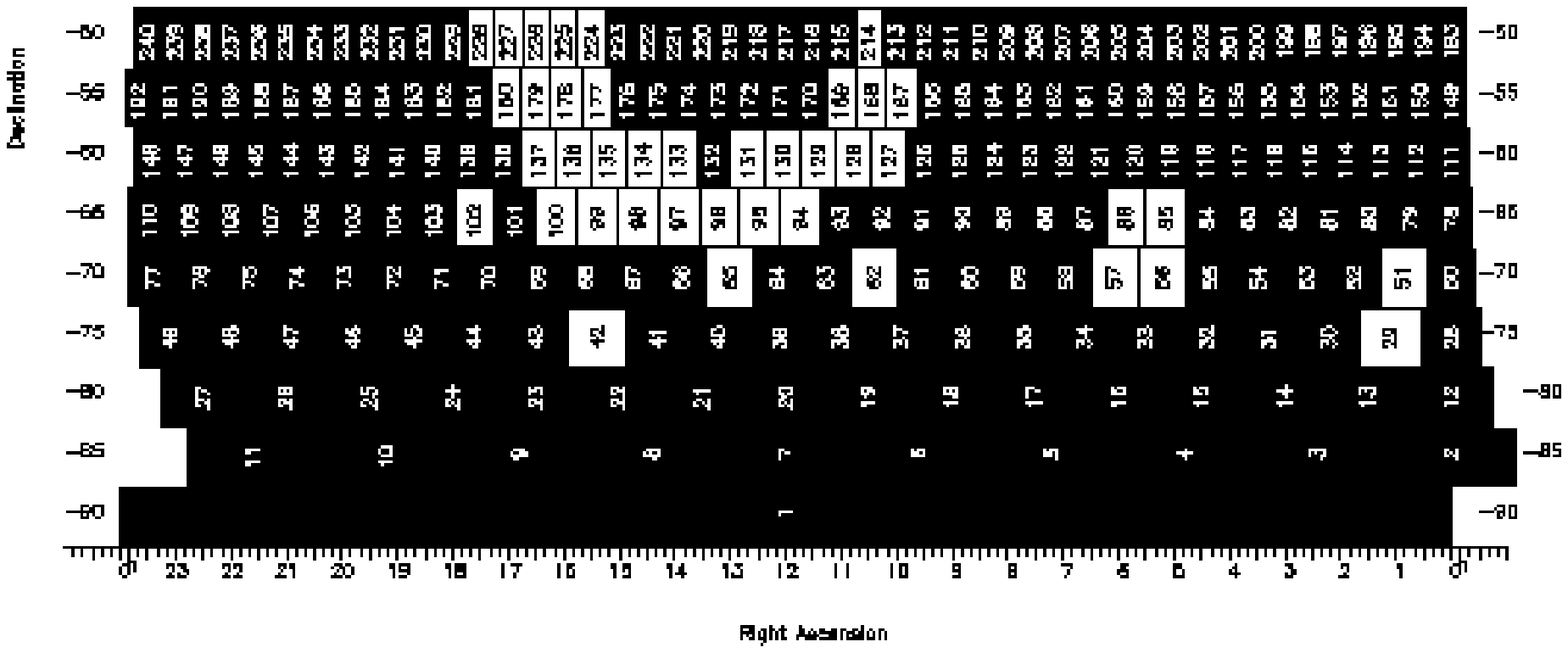]{Plate coverage of the sky included in the
  MINIMO search.  White boxes indicate plates that were excluded,
  primarily due to crowding in the Galactic Plane, LMC, and SMC, or
  because of an insufficient spread in plate epochs.
\label{platecoverage}}

\figcaption[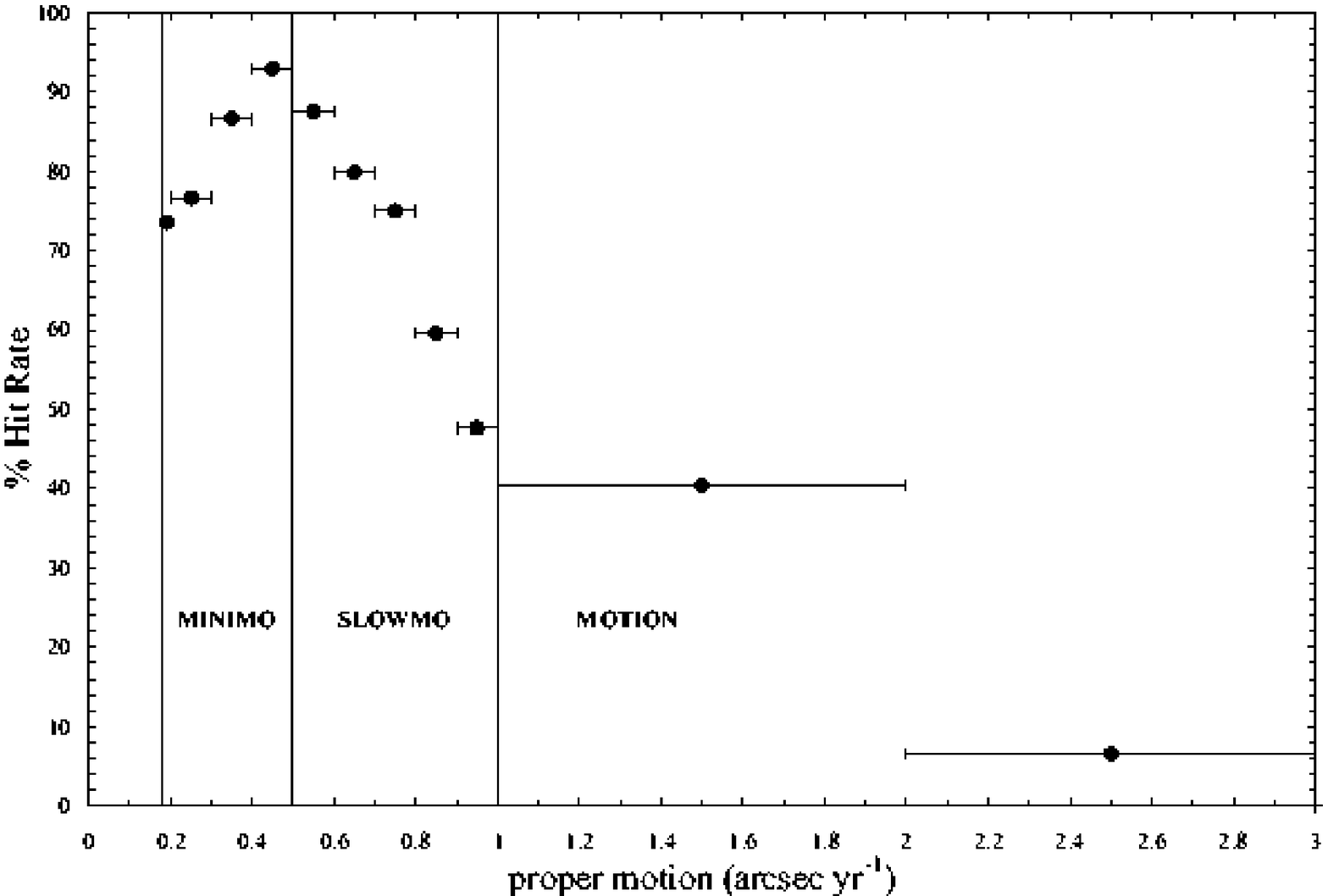]{Hit Rate percentage of true proper motion
  objects (real proper motion objects / total candidates extracted) for
  the entire SCR sample having $\mu$ $\ge$ 0$\farcs$18 yr$^{-1}$ south
  of $-$47$\degr$.  Horizontal bars show the proper motion bins
  adopted, while vertical lines delineate the three individual SCR
  samples --- MOTION, SLOWMO and MINIMO.
\label{hitrate}}

\figcaption[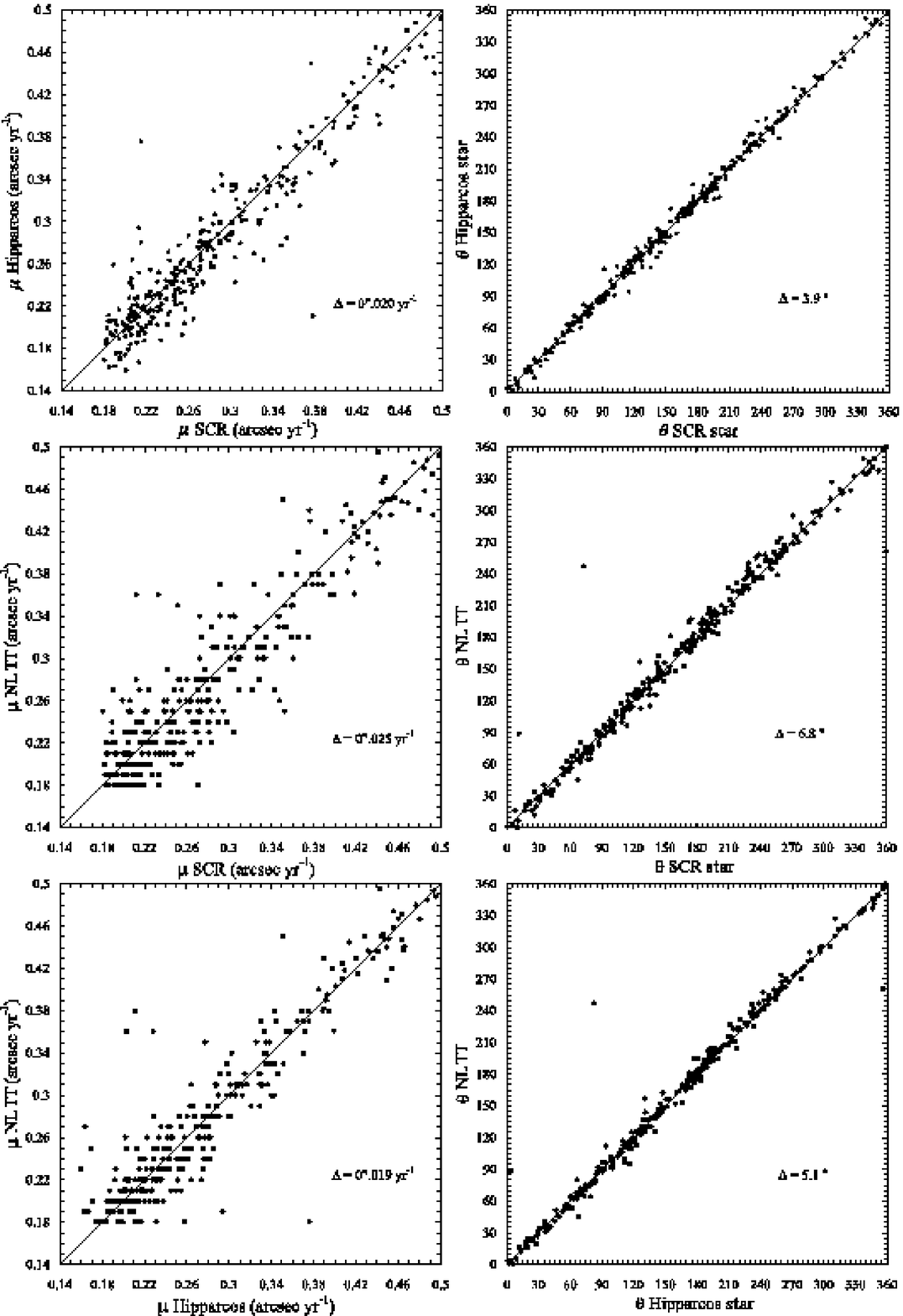]{Comparison of NLTT, Hipparcos, and SCR
  proper motions and position angles.  The solid line
  indicates perfect agreement.  Note that proper motion data for NLTT
  is quantized south of $-$45$\degr$.
\label{pm-error}}

\figcaption[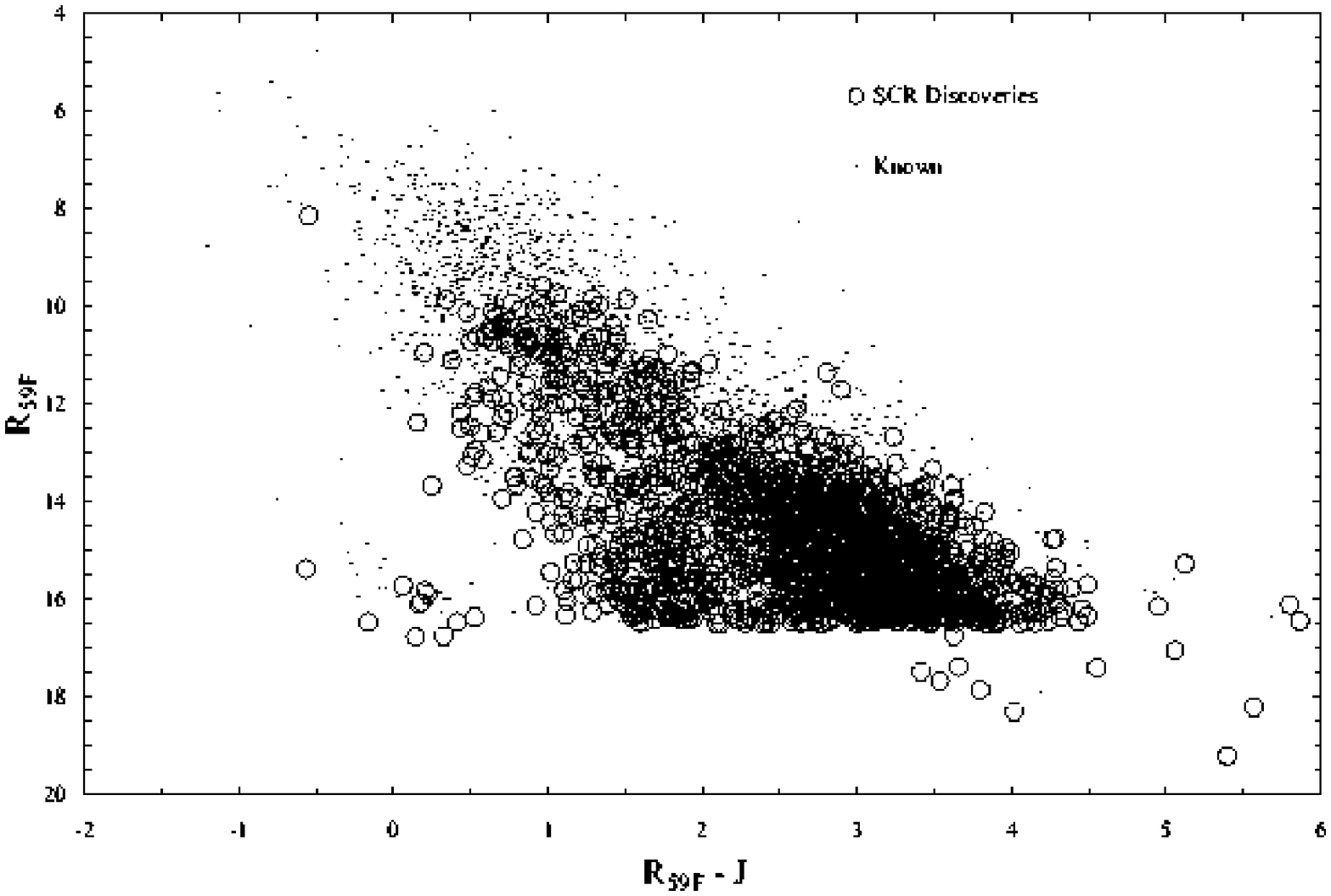]{Color-apparent magnitude diagram for SCR
  systems (open circles) and known systems (small points) with
  0$\farcs$40 yr$^{-1}$ $>$ $\mu$ $\ge$ 0$\farcs$18 yr$^{-1}$ from
  $-$90$\degr$ $>$ $\delta$ $\ge$ $-$47$\degr$.  Data points below
  $R_{59F}-J =$ 16.5 are CPM companions noticed during the blinking
  process.
\label{coloram}}

\figcaption[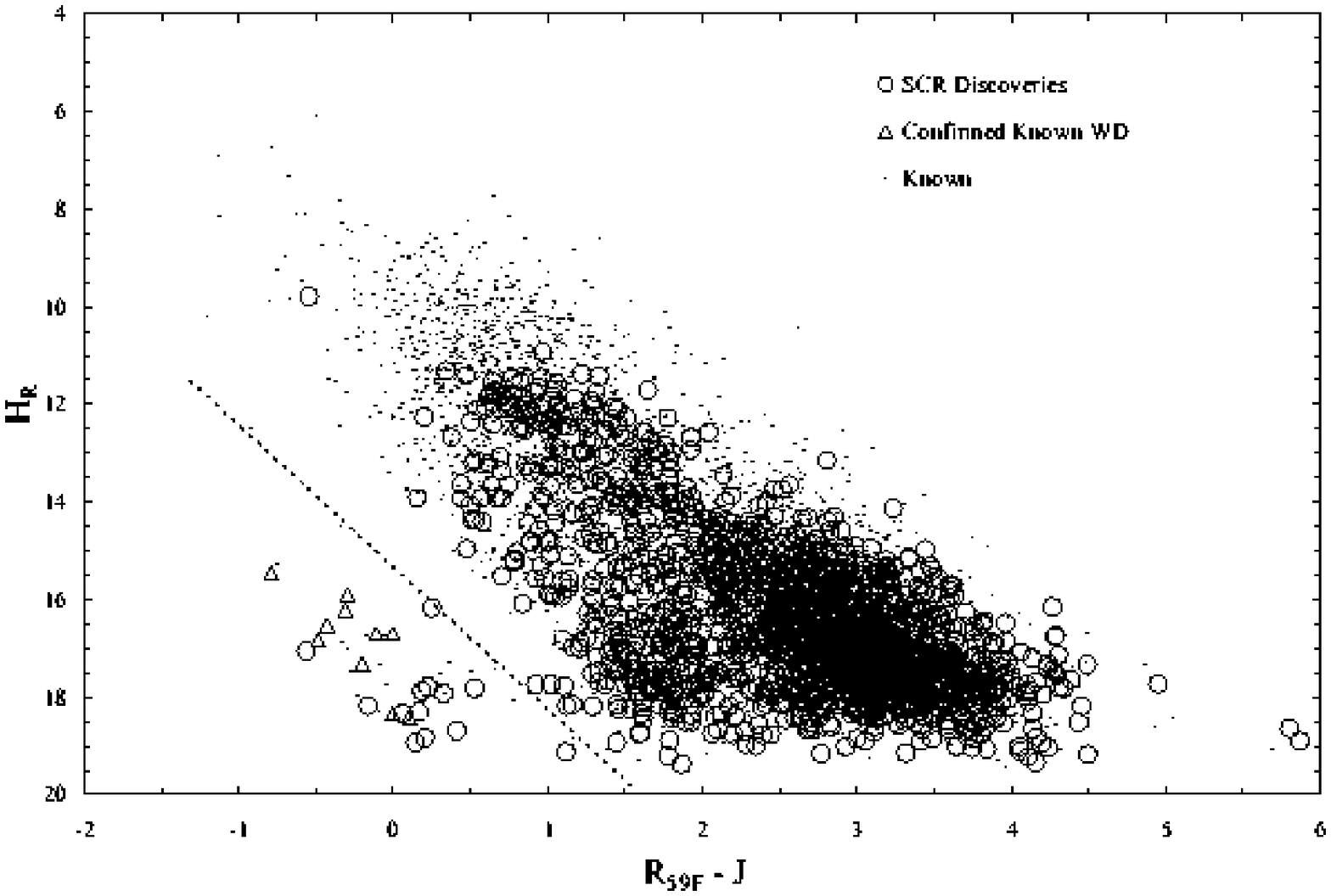]{Reduced proper motion diagram for SCR
  systems (open circles) and known systems (small points) with
  0$\farcs$40 yr$^{-1}$ $>$ $\mu$ $\ge$ 0$\farcs$18 yr$^{-1}$ from
  $-$90$\degr$ $>$ $\delta$ $\ge$ $-$47$\degr$.  The arbitrary dotted
  line separates the white dwarfs from the subdwarfs.  Triangles
  denote previously known white dwarfs.
\label{reducedpm}}

\figcaption[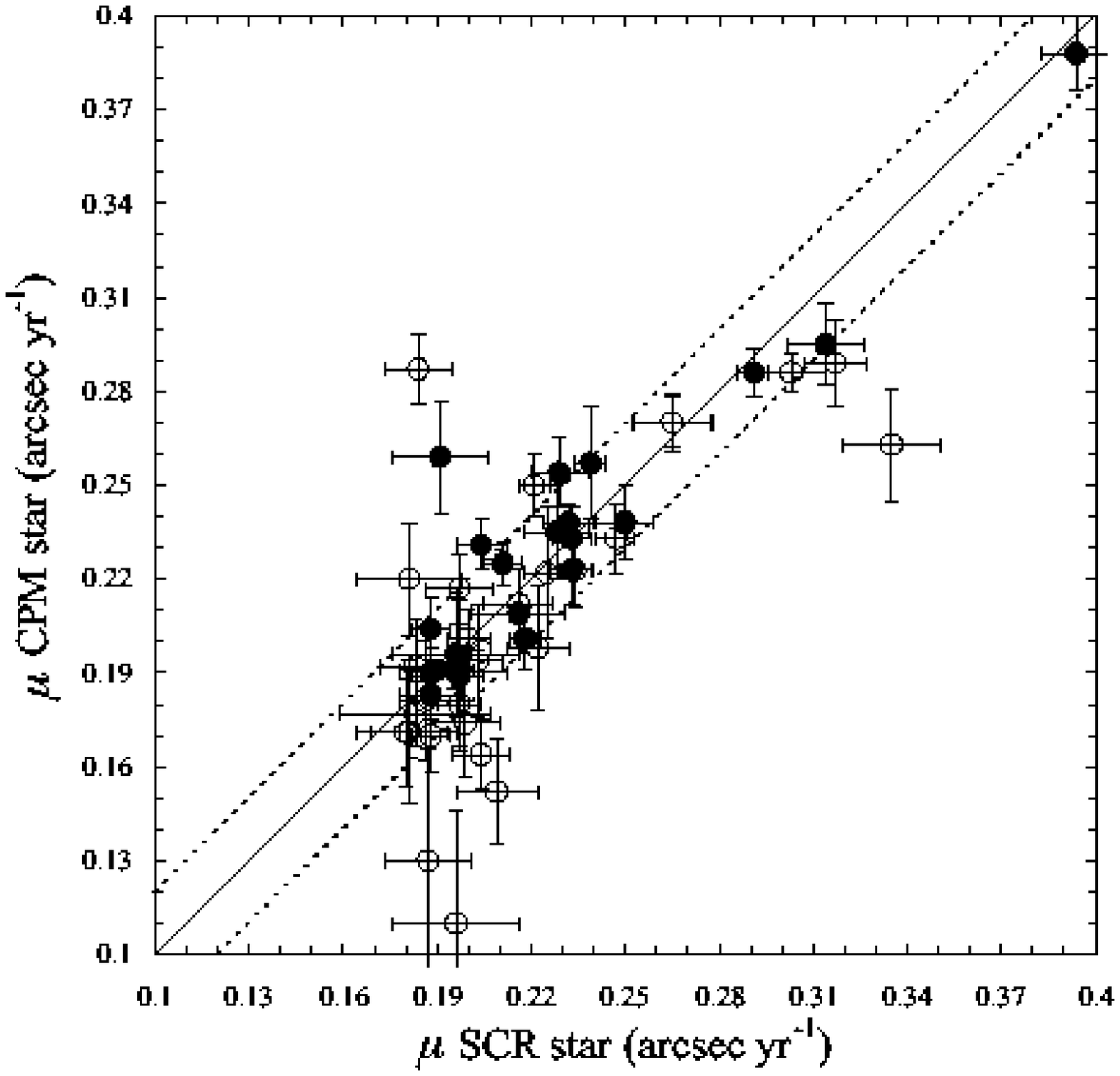]{Comparison of proper motions for components in CPM
  systems.  Proper motions from the automated SCR search are denoted
  by filled circles and proper motions manually obtained through
  SuperCOSMOS are denoted by open circles.  The solid line indicates
  perfect agreement between the two and the dashed lines represent
  conservative limits of 0$\farcs$020 yr$^{-1}$ based on our
  uncertanties.
\label{CPM-mu}}

\figcaption[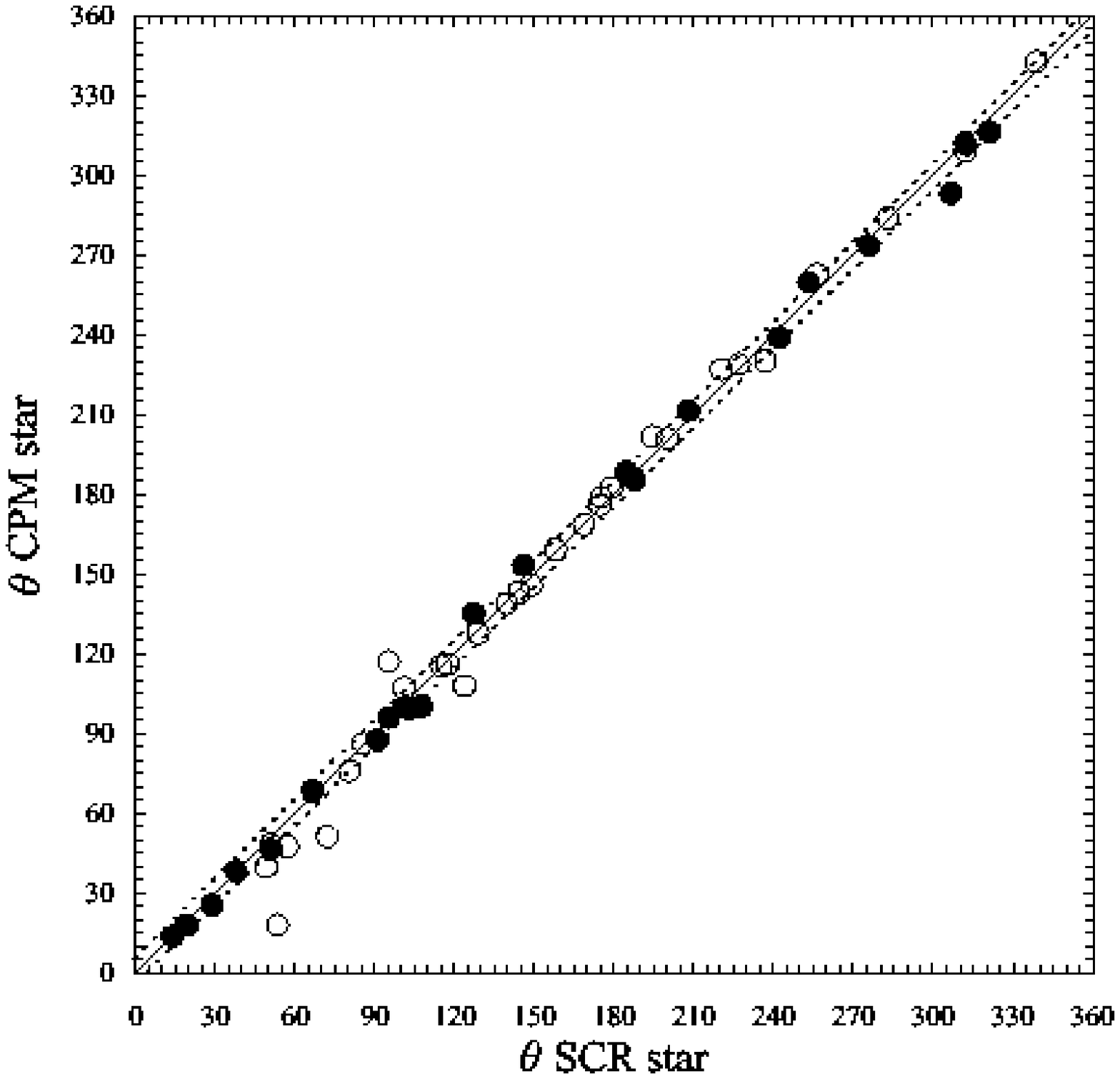]{Comparison of position angles for components in
  multiple CPM systems.  Position angles from the automated SCR search
  are denoted by filled circles and position angles manually obtained
  through SuperCOSMOS are denoted by open circles.  The solid line
  indicates perfect agreement between the two and the dashed lines
  represent conservative limits of 5.0 degrees based on our uncertanties.
\label{CPM-pa}}

\figcaption[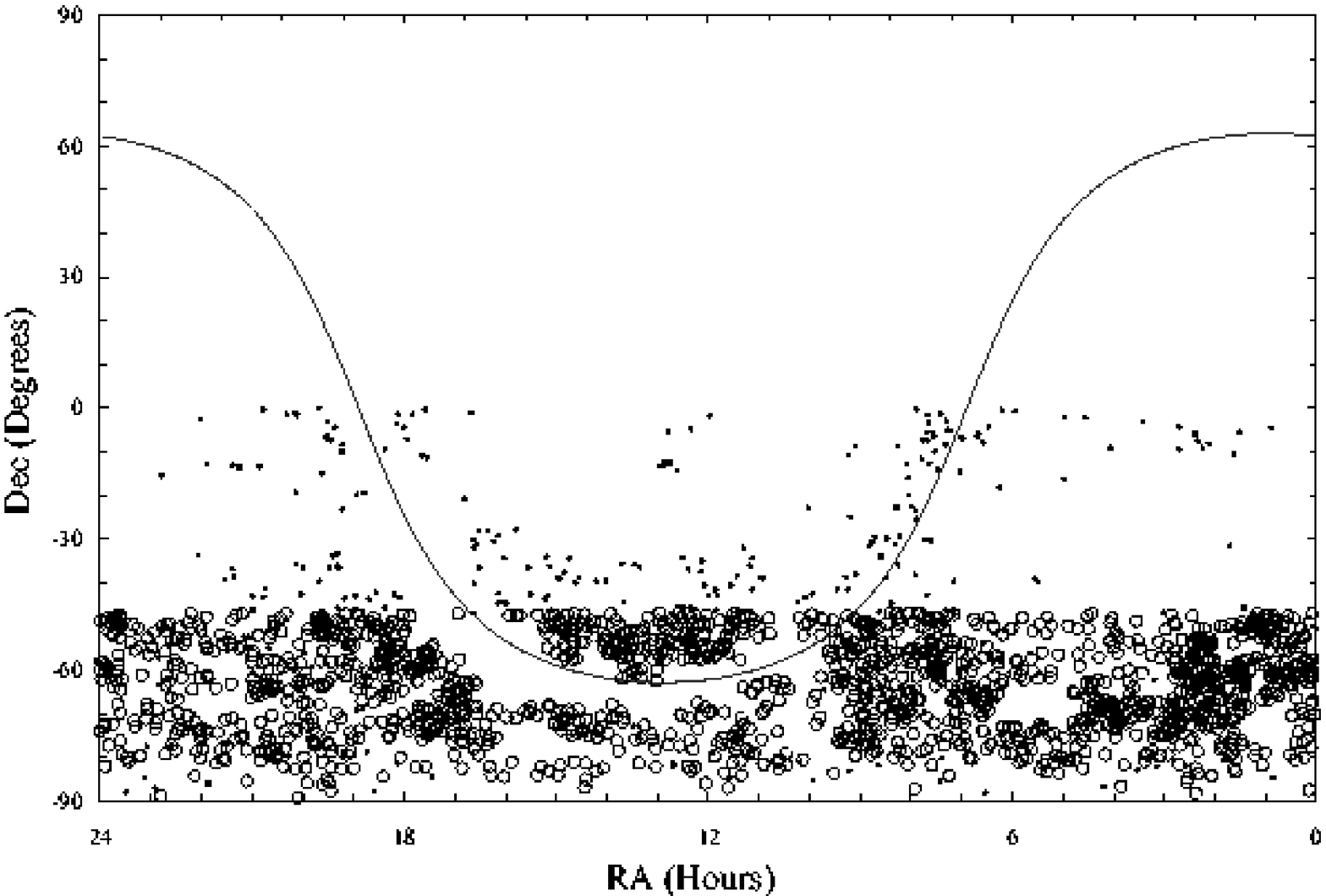]{Sky distribution of SCR systems with $\mu$
  $\ge$ 0$\farcs$18 yr$^{-1}$.  Open circles denote SCR discoveries
  from this paper and filled circles indicate SCR discoveries from
  past papers.  The curve represents the Galactic plane.
\label{skycoverage}}

\clearpage


 \begin{figure}
 \plotone{f1.ps}
 \end{figure}
 
 \clearpage
 
 \begin{figure}
 \plotone{f2.ps}
 \end{figure}
 
 \clearpage
 
 \begin{figure}
 \plotone{f3.ps}
 \end{figure}
 
 \clearpage
 
 \begin{figure}
 \plotone{f4.ps}
 \end{figure}
 
 \clearpage
 
 \begin{figure}
 \plotone{f5.ps}
 \end{figure}
 
 \clearpage
 
 \begin{figure}
 \plotone{f6.ps}
 \end{figure}
 
 \clearpage
 
 \begin{figure}
 \plotone{f7.ps}
 \end{figure}
 
  \clearpage

 \begin{figure}
 \plotone{f8.ps}
 \end{figure}


\begin{thebibliography}{}

\bibitem[Deacon et al.(2005)]{2005A&A...435..363D} Deacon, N.~R.,
  Hambly, N.~C., \& Cooke, J.~A.\ 2005, \aap, 435, 363

\bibitem[Giclas et al.(1971)]{1971lpms.book.....G} Giclas, H.~L.,
  Burnham, R., \& Thomas, N.~G.\ 1971, Flagstaff, Arizona: Lowell
  Observatory, 1971

\bibitem[Giclas et al.(1978)]{1978LowOB...8...89G} Giclas, H.~L.,
  Burnham, R., \& Thomas, N.~G.\ 1978, Lowell Observatory Bulletin, 8,
  89

\bibitem[Hambly et al.(2004)]{2004AJ....128..437H} Hambly, N.~C.,
  Henry, T.~J., Subasavage, J.~P., Brown, M.~A., \& Jao, W.\ 2004,
  \aj, 128, 437

\bibitem[Hambly et al.(2001a)]{2001MNRAS.326.1315H} Hambly, N.~C.,
  Davenhall, A.~C., Irwin, M.~J., \& MacGillivray, H.~T.\ 2001c,
  \mnras, 326, 1315

\bibitem[Hambly et al.(2001b)]{2001A&A....326..1295H} Hambly, N.~C.,
  Irwin, M.~J., \& MacGillivray, H.~T.\ 2001, \mnras, 326, 1295

\bibitem[Henry et al.(2006)]{2006AJ....132.2360H} Henry, T.~J., Jao,
  W.-C., Subasavage, J.~P., Beaulieu, T.~D., Ianna, P.~A., Costa, E.,
  \& M{\'e}ndez, R.~A.\ 2006, \aj, 132, 2360

\bibitem[Henry et al.(2004)]{2004AJ....128.2460H} Henry, T.~J.,
  Subasavage, J.~P., Brown, M.~A., Beaulieu, T.~D., Jao, W.-C., \&
  Hambly, N.~C.\ 2004, \aj, 128, 2460

\bibitem[Jao(2004)]{2004PhDT........11J} Jao, W.-C.\ 2004,
  Ph.D.~Thesis

\bibitem[Jao et al.(2005)]{2005AJ....129.1954J} Jao, W.-C., Henry,
  T.~J., Subasavage, J.~P., Brown, M.~A., Ianna, P.~A., Bartlett,
  J.~L., Costa, E., \& M{\'e}ndez, R.~A.\ 2005, \aj, 129, 1954

\bibitem[L{\'e}pine \& Shara(2005)]{2005AJ....129.1483L} L{\'e}pine,
  S., \& Shara, M.~M.\ 2005, \aj, 129, 1483

\bibitem[L{\'e}pine(2005)]{2005AJ....130.1247L} L{\'e}pine, S.\ 2005,
  \aj, 130, 1247

\bibitem[Luyten(1979)]{1979lccs.book.....L} Luyten, W.~J.\ 1979, LHS
  Catalogue (Minneapolis: Univ. of Minnesota Press)

\bibitem[Luyten(1980)]{1980PMMin..55....1L} Luyten, W.~J.\ 1980,
  Proper Motion Survey with the 48-inch Telescope, Univ.~Minnesota,
  55, 1 (1980), 55, 1

\bibitem[Luyten(1995)]{1995yCat.1098....0L} Luyten, W.~J.\ 1995,
  VizieR Online Data Catalog, 1098, 0

\bibitem[Oppenheimer et al.(2001)]{2001Sci...292..698O} Oppenheimer,
  B.~R., Hambly, N.~C., Digby, A.~P., Hodgkin, S.~T., \& Saumon,
  D.\ 2001, Science, 292, 698

\bibitem[Pokorny et al.(2003)]{2003A&A...397..575P} Pokorny, R.~S.,
  Jones, H.~R.~A., \& Hambly, N.~C.\ 2003, \aap, 397, 575

\bibitem[Pokorny et al.(2004)]{2004A&A...421..763P} Pokorny, R.~S.,
  Jones, H.~R.~A., Hambly, N.~C., \& Pinfield, D.~J.\ 2004, \aap, 421,
  763

\bibitem[Ruiz et al.(1993)]{1993AJ....106.2575R} Ruiz, M.~T.,
  Takamiya, M.~Y., Mendez, R., Maza, J., \& Wishnjewsky, M.\ 1993,
  \aj, 106, 2575

\bibitem[Ruiz et al.(2001)]{2001ApJS..133..119R} Ruiz, M.~T.,
  Wischnjewsky, M., Rojo, P.~M., \& Gonzalez, L.~E.\ 2001, \apjs, 133,
  119

\bibitem[Scholz et al.(2000)]{2000A&A...353..958S} Scholz, R.-D.,
  Irwin, M., Ibata, R., Jahrei{\ss}, H., \& Malkov, O.~Y.\ 2000, \aap,
  353, 958

\bibitem[Scholz et al.(2002)]{2002ApJ...565..539S} Scholz, R.-D.,
  Szokoly, G.~P., Andersen, M., Ibata, R., \& Irwin, M.~J.\ 2002,
  \apj, 565, 539

\bibitem[Subasavage et al.(2005a)]{2005AJ....129..413S} Subasavage,
  J.~P., Henry, T.~J., Hambly, N.~C., Brown, M.~A., \& Jao,
  W.~C.\ 2005, \aj, 129, 413

\bibitem[Subasavage et al.(2005b)]{2005AJ....130.1658S} Subasavage,
  J.~P., Henry, T.~J., Hambly, N.~C., Brown, M.~A., Jao, W.~C., \&
  Finch, C.~T.\ 2005, \aj, 130, 1658

\bibitem[Wroblewski \& Costa(1999)]{1999A&AS..139...25W} Wroblewski,
  H., \& Costa, E.\ 1999, \aaps, 139, 25

\bibitem[Wroblewski \& Costa(2001)]{2001A&A...367..725W} Wroblewski,
  H., \& Costa, E.\ 2001, \aap, 367, 725

\bibitem[Wroblewski \& Torres(1989)]{1989A&AS...78..231W} Wroblewski,
  H., \& Torres, C.\ 1989, \aaps, 78, 231

\bibitem[Wroblewski \& Torres(1991)]{1991A&AS...91..129W} Wroblewski,
  H., \& Torres, C.\ 1991, \aaps, 91, 129

\bibitem[Wroblewski \& Torres(1994)]{1994A&AS..105..179W} Wroblewski,
  H., \& Torres, C.\ 1994, \aaps, 105, 179

\bibitem[Wroblewski \& Torres(1996)]{1996A&AS..115..481W} Wroblewski,
  H., \& Torres, C.\ 1996, \aaps, 115, 481

\bibitem[Wroblewski \& Torres(1997)]{1997A&AS..122..447W} Wroblewski,
  H., \& Torres, C.\ 1997, \aaps, 122, 447

\end{thebibliography}
\end{document}